\documentclass[twocolumn, aps, prl, 10pt, floatfix, superscriptaddress]{revtex4-1}

\usepackage{lmodern}
\usepackage[T1]{fontenc}
\usepackage[utf8]{inputenc}
\usepackage[english]{babel}

\usepackage{amsmath}
\usepackage{amssymb}
\usepackage{mathtools}
\usepackage{bm}
\usepackage{graphicx,color}
\usepackage{hyperref}
\usepackage{booktabs}
\usepackage{multirow}
\usepackage{epstopdf}

\hypersetup{
  colorlinks   = true, 
  urlcolor     = blue, 
  linkcolor    = blue, 
  citecolor   = red, 
}

\newcommand{\tr}{\operatorname{tr}}
\newcommand{\ic}{\ensuremath{\mathrm{i}}}

\newcommand{\ket}[1]{|#1\rangle}
\newcommand{\bra}[1]{\langle#1|}

\def\eye{I}
\def\CFT{\textsl{\tiny CFT}}
\def\TCI{\textsl{\tiny TCI}}
\def\QFT{\textsl{\tiny QFT}}

\begin{document}

\title{Conformal data and renormalization group flow in critical quantum spin chains \\ using periodic uniform matrix product states}


\author{Yijian Zou}
\email[]{yzou@pitp.ca}
\affiliation{Perimeter Institute for Theoretical Physics, Waterloo ON, N2L 2Y5, Canada}
\affiliation{University of Waterloo, Waterloo ON, N2L 3G1, Canada}
\author{Ashley Milsted}
\email[]{amilsted@pitp.ca}
\affiliation{Perimeter Institute for Theoretical Physics, Waterloo ON, N2L 2Y5, Canada}
\author{Guifre Vidal}
\affiliation{Perimeter Institute for Theoretical Physics, Waterloo ON, N2L 2Y5, Canada}

\date{\today}

\begin{abstract}
We establish that a Bloch-state ansatz based on periodic uniform Matrix Product States (puMPS), originally designed to capture single-quasiparticle excitations in gapped systems, is in fact capable of accurately approximating \textit{all}
low-energy eigenstates of critical quantum spin chains on the circle. 
When combined with the methods of [Milsted, Vidal, Phys. Rev. B 96 245105] based on the Koo-Saleur formula, puMPS Bloch states can then be used to identify each low-energy eigenstate of a chain made of up to hundreds of spins with its corresponding scaling operator in the emergent conformal field theory (CFT). This enables the following two tasks, that we  
demonstrate using the quantum Ising model and a recently proposed generalization thereof due to O'Brien and Fendley [Phys. Rev. Lett. 120, 206403]. (i) From the spectrum of low energies and momenta we extract conformal data (specifying the emergent CFT) with unprecedented numerical accuracy. (ii) By changing the lattice size, we investigate nonperturbatively the RG flow of the low-energy spectrum between two CFTs. In our example, where the flow is from the Tri-Critical Ising CFT to the Ising CFT, we obtain excellent agreement with an analytical result [Klassen and Melzer, Nucl. Phys. B 370 511] conjectured to describe the flow of the first spectral gap directly in the continuum.
\end{abstract}

\maketitle

Near a continuous phase transisiton, two microscopically different systems are assigned to the same universality class if they display similar long-distance behavior \cite{wilson_renormalization_1974}.
In the language of the renormalization group (RG), which describes how physics changes with scale, such systems are said to ``flow'' to the same scale-invariant theory or RG fixed point. These fixed points are often described by a conformal field theory (CFT) \cite{belavin_infinite_1984,friedan_conformal_1984}, which itself is specified by a set of parameters known as conformal data.
 
Given a microscopic description of a critical system (e.g.\ a lattice Hamiltonian), an important, yet challenging task is to extract the conformal data as a means of identifying the universality class of the phase transition.  For critical quantum spin chains -- the focus of this work -- much progress can be made in highly fine-tuned models, such as integrable lattice models (for example, \cite{zamolodchikov_thermodynamic_1991, feverati_lattice_2002, pearce_excited_2003, read_enlarged_2007, gainutdinov_lattice_2013, bondesan_chiral_2015,  zini_conformal_2017}). However, for a generic critical spin chain Hamiltonian one must resort to numerical methods. Exact diagonalization techniques are certainly useful \cite{feiguin_interacting_2007, milsted_extraction_2017}, but can only address small systems, where the universal low-energy physics is often concealed by the non-universal, microscopic details. Monte Carlo methods can address much larger systems \cite{sandvik_computational_2010-1}, but only in models that do not suffer from the sign problem. On the other hand, tensor network methods \cite{white_1992, fannes_1992, vidal_2007a} are both sign-problem free and scalable, and several schemes have been proposed to extract conformal data \cite{degli_investigation_2004, tagliacozzo_2008, xavier_entanglement_2010, evenbly_quantum_2013, stojevic_2015, evenbly_tensor_2015}. These include schemes \cite{degli_investigation_2004, tagliacozzo_2008, xavier_entanglement_2010, stojevic_2015} based on the matrix product state (MPS) \cite{fannes_1992, vidal_2004}, which is the ansatz underlying the density matrix renormalization group (DMRG) algorithm \cite{white_1992, schollwock_2011}.

In this Letter we establish that a Bloch-state ansatz~\cite{rommer_1997} based on periodic uniform Matrix Product States (puMPS) \cite{pirvu_exploiting_2011} is ideally suited to  numerically investigate the emergent universal properties of critical quantum spin chains (see Ref.~\cite{pirvu_matrix_2012} for previous use in critical systems). Our key observation is that, despite being originally designed to capture only \textit{some} (namely single-quasiparticle) low-energy excitations in \textit{gapped} systems \cite{rommer_1997}, puMPS Bloch states turn out to accurately reproduce \emph{all} low-energy eigenstates of \textit{critical} quantum spin chains (that is, up to some appropriate maximum energy) \footnote{This observation had been made for two integrable models \cite{pirvu_exploiting_2011}. We establish its validity for generic critical quantum spin chains via several examples, both integrable and nonintegrable \cite{supplemental}.}. The ability of puMPS to simulate systems consisting of several hundreds of spins allow us to then put forward two new applications of this tensor network ansatz: (i) extraction, with unprecedented accuracy, of the conformal data characterizing the underlying CFT and thus the universality class of the corresponding continuous phase transition; (ii) nonperturbative computation of the RG flow of the low-energy spectrum between two CFTs. Here we demonstrate these applications using the quantum Ising model and its recently proposed generalization due to O'Brien and Fendley \cite{obrien_lattice_2018}, with which we study the spectral RG flow between the Tri-Critical Ising CFT and the Ising CFT. We find excellent agreement between our numerical results and an analytical result \cite{klassen_spectral_1992} conjectured to describe the flow of the first spectral gap directly in the continuum.

\emph{Matrix Product State ansatz.} Given a local Hamiltonian $H$ for a critical quantum spin chain of $N$ spins on the circle, we compute approximations to the ground state and excited states. For the ground state we use a puMPS $|\Psi(A)\rangle$ \cite{pirvu_exploiting_2011}, which is specified by a tensor $A^s_{ab}$ of dimension $d \times D \times D$, where $d$ is the dimension of the Hilbert space of one spin and $D$ is the \emph{bond dimension}, which restricts the amount of \emph{entanglement}. It has the translation-invariant form $|\Psi(A)\rangle \equiv \sum_{\vec{s}=1}^d \tr \left( A^{s_1} A^{s_2} \dots A^{s_N} \right) |\vec{s}\,\rangle$, where $\vec{s} = s_1\dots s_N$. To find the variational ground state, we minimize the energy with respect to $A^s_{ab}$ using a gradient descent method \cite{ganahl_continuous_2016, supplemental}.
We then seek excitations within the space of puMPS \emph{Bloch states} \cite{rommer_1997, pirvu_matrix_2012}, which have the form
\begin{equation} \label{puMPS_ex}
  |\Phi_p(B)\rangle \equiv \sum_{j=1}^N e^{-\ic p j} \mathcal{T}^j \sum_{\vec{s}=1}^d \tr \left( B^{s_1} A^{s_2} \dots A^{s_N} \right) |\vec{s}\,\rangle,
\end{equation}
where $\mathcal{T}$ is the translation operator, $p$ is the momentum, $A$ is the ground-state puMPS tensor, and $B^{s}_{ab}$ is a tensor that parameterizes a Bloch state and is obtained by diagonalizing an effective Hamiltonian \cite{supplemental}.

We assess the performance of the ansatz \eqref{puMPS_ex} for critical systems using the critical transverse field Ising model $H = -\sum_{j=1}^N \left[ \sigma^X_j\sigma^X_{j+1} + \sigma^Z_j \right]$. As a first test, we compute the variational ground state and excitations $\ket{\phi_{\alpha}}$ for a small system of $N=20$ spins and check the fidelity $f_{\alpha} \equiv \bra{\phi_{\alpha}}{\phi_{\alpha}^{\mbox{\tiny exact}}}\rangle$ with their counterparts computed using exact diagonalization. We find, for fixed bond-dimension $D=12$, that the first 41 excited states have errors $\epsilon_{\alpha} \equiv 1-|f_{\alpha}|^2$ ranging from $~10^{-4}$ to $~10^{-11}$,
and that this error always scales to zero with increasing $D$.
To test larger systems, where we can no longer use exact diagonalization, we compare the low-energy spectrum of excitation energies with the CFT prediction for the $N\rightarrow \infty$ limit. We find that all variational low-energy excitations have energies consistent with the CFT prediction up to a maximum energy that depends on the system size $N$ and the bond dimension $D$, see \cite{supplemental} for more details. We conclude that \textit{all} low-energy excitations are well-approximated by the Bloch-state puMPS \eqref{puMPS_ex}. This is remarkable, given that this ansatz was originally proposed \cite{rommer_1997} for \textit{single-quasiparticle} excitations in \emph{gapped} systems, where multi-quasiparticle excitations require an alternative, significantly more sophisticated ansatz \cite{vanderstraeten_2015}.

\emph{Extracting conformal data.}
Given excited states of the critical spin chain, we wish to extract conformal data of the 2D CFT describing its RG fixed point. This includes the \textit{central charge} $c$ and the \textit{scaling dimensions} $\Delta_\alpha$ and \textit{conformal spins} $S_\alpha$ of a subset of \emph{scaling operators} $\phi_{\alpha}$ (CFT operators that are covariant under dilations and rotations), namely those known as \textit{primary fields} \cite{belavin_infinite_1984}.
There are several useful results \cite{cardy_conformal_1984, blote_1986, affleck_universal_1986, cardy_operator_1986, koo_representations_1994, holzhey_geometric_1994, calabrese_entanglement_2004, alcaraz_entanglement_2011} that relate quantities computed from a finite spin chain to this conformal data. Here we make use of the discovery \cite{cardy_conformal_1984, blote_1986, affleck_universal_1986, cardy_operator_1986} that the eigenstates of $H$ have energies $E_\alpha$ and momenta $P_\alpha$ given by
\begin{equation} \label{circle_spec}
E_\alpha=A+B\frac{2\pi}{N}(\Delta_\alpha-\frac{c}{12})+O(N^{-x}), \quad
P_\alpha=\frac{2\pi}{N}S_\alpha,
\end{equation}
where $N$ is the number of spins, and $A$, $B$, $x$ are constants specific to the microscopic model $H$, with $x > 1$ determining subleading corrections to the dominant scaling with $N$. 
Up to these constants, \eqref{circle_spec} is determined by universal quantities, with each pair $\Delta_\alpha, S_\alpha$ corresponding to a CFT scaling operator $\phi_{\alpha}$ via the \emph{operator-state correspondence} \cite{belavin_infinite_1984}. Indeed, we can identify each eigenstate $\ket{\phi_{\alpha}}$ with a CFT operator $\phi_{\alpha}$ using the methods of Ref.~\cite{milsted_extraction_2017} based on approximate lattice representations~\cite{koo_representations_1994}
\begin{equation} \label{Hn}
  H_n = \frac{N}{2\pi} \sum_{j=1}^N e^{\ic jn \frac{2\pi}{N}} h_j \quad \sim \quad L_n + \overline{L}_{-n},
\end{equation}
of the Virasoro generators $L_n,\overline{L}_n$ of conformal transformations \cite{belavin_infinite_1984}. These act as \emph{ladder operators} on the eigenstates $\ket{\phi_{\alpha}}$ of $H$, which are organized into \emph{conformal towers} of states, each descended from a distinct primary field state. To illustrate how the above identification $\ket{\phi_{\alpha}} \sim \phi_{\alpha}$ works, here are 3 examples: 
(i) Lattice energy eigenstates corresponding to CFT primary operators are those that can not be lowered in energy by any of $H_{\pm 1}$, $H_{\pm 2}$ (up to some matrix elements that decay with system size) \cite{milsted_extraction_2017}. For instance, in a unitary CFT the ground state of the critical spin chain is always identified with the primary identity operator $\eye$  \cite{belavin_infinite_1984}, hence it receives the label $|\eye\rangle$. 
(ii) The lattice state $|T\rangle$ corresponding to the stress tensor operator $T$ \cite{belavin_infinite_1984} is characterized as the energy eigenstate $\ket{\psi}$ which maximizes $|\langle \psi|H_{-2}|\eye\rangle|$, in analogy with the CFT relation $L_{-2}|\eye\rangle_\CFT = \sqrt{\frac{c}{2}}|T\rangle_\CFT$. Below we use eigenstates $\ket{\eye}$ and $|T\rangle$ to
compute an estimate of the central charge $c \approx 2 |\langle T|H_{-2}|\eye\rangle|^2$ \cite{koo_representations_1994}. 
(iii) The CFT analogue of $H_{2}H_{-2}|\eye\rangle$ is $(L_2 + \overline{L}_{-2})L_{-2}|\eye\rangle_\CFT = a|\eye\rangle_\CFT + b|T\overline{T}\rangle_\CFT$, where $a$ and $b$ are constants of order 1 determined by conformal symmetry, and we have used $\overline{L}_{2}|\eye\rangle_\CFT = 0$. We may thus identify the lattice state $|T\overline{T}\rangle$ corresponding to the operator $T\overline{T}$ \cite{mcgough_moving_2016, cavaglia_t_2016, smirnov_space_2017} as the energy eigenstate $|\psi\rangle \not = \ket{I}$ that maximizes $|\langle \psi|H_{2}H_{-2}|\eye\rangle|$. 

\begin{table}[h]
  \begin{center}
    \footnotesize
    \begin{tabular}{llll}
      \multicolumn{4}{c}{Critical Ising model}\\
\toprule
 & exact & puMPS & error \\
\midrule
$c$ & 0.5  & 0.4999997 & $~10^{-7}$ \\
$\Delta_\sigma$ & 0.125 & 0.1249995 & $~10^{-7}$ \\
$\Delta_\varepsilon$ & 1 & 0.9999994 & $~10^{-7}$ \\
$\Delta_{\partial\bar{\partial}\sigma}$ & 2.125 & 2.12501 & $~10^{-5}$ \\
$\Delta_{\partial\bar{\partial}\varepsilon}$ & 3 & 3.00002 & $~10^{-5}$ \\
$\Delta_{T\bar{T}}$ & 4 & 4.007 & $~10^{-3} $ \\
\bottomrule
  \end{tabular}
  \hspace{0.2cm}
  \begin{tabular}{llll}
    \multicolumn{4}{c}{OF model, TCI point}\\
\toprule
 & exact & puMPS & error \\
\midrule
$c$ & 0.7  & 0.6991 & $~10^{-4}$ \\
$\Delta_\sigma$ & 0.075 & 0.07492 & $~10^{-5}$ \\
$\Delta_\varepsilon$ & 0.2 & 0.2001 & $~10^{-4}$ \\
$\Delta_{\sigma'}$ & 0.875 & 0.8747 & $~10^{-4}$ \\
$\Delta_{\varepsilon'}$ & 1.2 & 1.203 & $~10^{-3}$ \\
$\Delta_{\varepsilon''}$ & 3.0 & 3.002 & $~10^{-3}$ \\
\bottomrule
\end{tabular}
\end{center}
\caption{\label{tab:extrap} Central charge and selected scaling dimensions from lattice Virasoro matrix elements \cite{milsted_extraction_2017} and energy gaps derived from puMPS Bloch states \cite{supplemental}. For the Ising model, we used system sizes $N\le 228$ and bond dimensions $24\le D \le 49$. For the OF model near its Tri-Critical Ising (TCI) point, we used $N \le 128$ and $28 \le D \le 44$ (requiring more computational time than used for the Ising model \cite{supplemental}). Note the good agreement in the latter case, despite being slightly off-critical.}
\end{table}

For instance, for $N=64$ spins and bond dimension $D=24$, we obtain a correct identification of all low-energy states $\ket{\phi_{\alpha}}$ of the critical Ising model with scaling operators $\phi_{\alpha}$ of the Ising CFT up to scaling dimension $\Delta_{\alpha} = 6$ (see~\cite{supplemental} for plots). We then compute variational excitations for a number of system sizes $N$ and, by extrapolating to large $N$, we estimate the scaling dimensions of a selection of scaling operators, as well as the central charge using \eqref{Hn}: see Table~\ref{tab:extrap}. We obtain excellent accuracy, with our results being consistently better than those from other methods, such as finite-entanglement scaling with infinite MPS \cite{stojevic_2015}, or MERA and TNR techniques \cite{evenbly_quantum_2013, hauru_topological_2016} (see \cite{supplemental} for a detailed comparison). The computations for this test were carried out in a matter of minutes on a modestly powerful laptop. In addition, the algorithms we use, despite being somewhat more complicated than DMRG, are significantly simpler than those required for the aforementioned methods.

\begin{figure}
  \includegraphics[width=\linewidth]{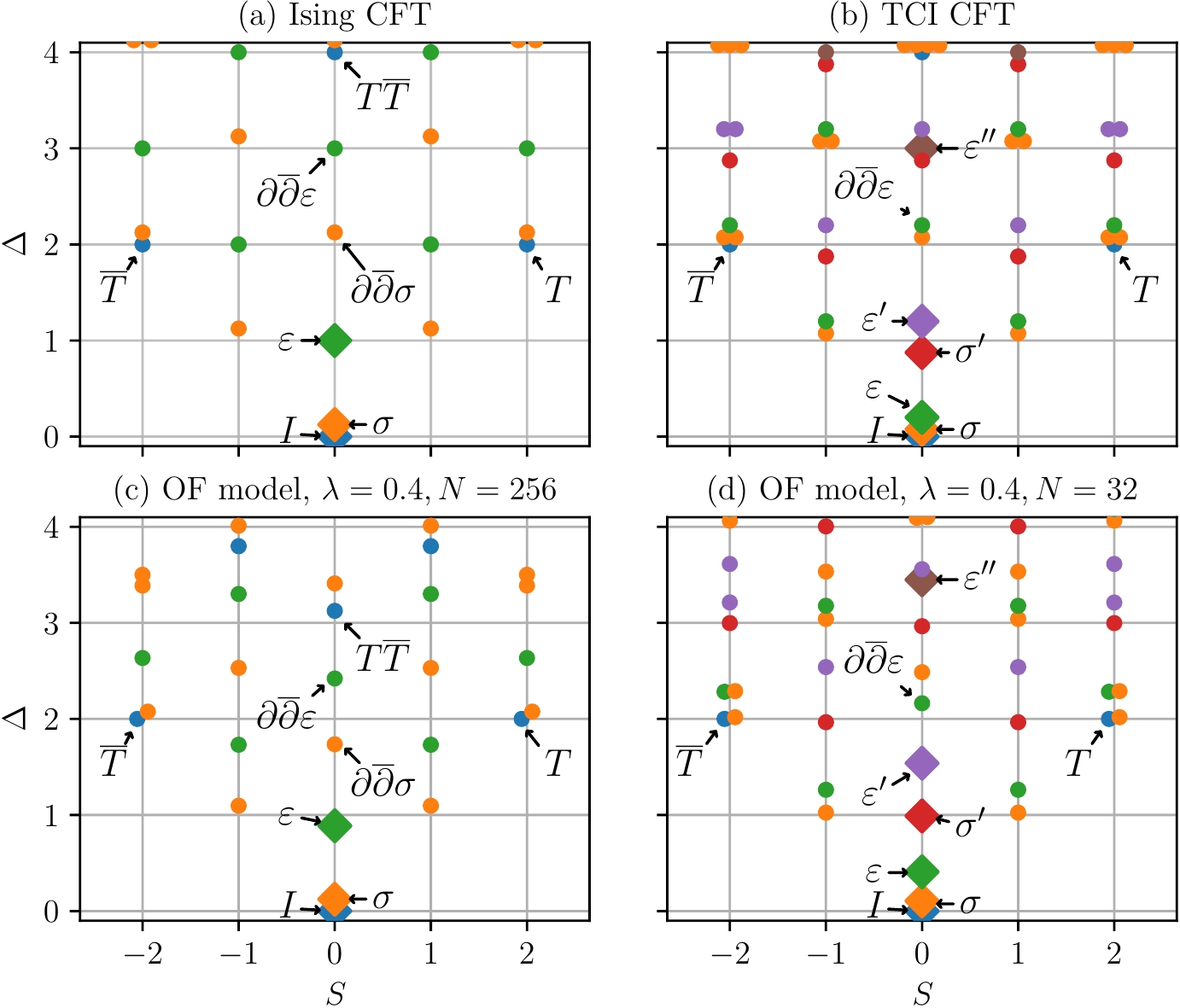}\\
  \vspace{-0.8em}
  \caption{\label{fig:4specs} {\bf Top:}\ Scaling operator spectra of (a) the Ising and (b) the TCI CFTs (with a selection of operators labeled). {\bf Bottom:}\ Approximate scaling dimensions and conformal-tower identification for the OF model at $\lambda=0.4$ with (c) $N=256, D=52$ and (d) $N=32, D=32$, corresponding to points of Fig.~\ref{fig:flow_N}. We label a selection of states according to a numerical identification of the corresponding CFT operators~\cite{milsted_extraction_2017, supplemental}. {\bf Note:}\ We displace data points slightly along the x-axis to show degeneracies.}
\end{figure}

\emph{Spectral RG flow.} We now turn \footnote{We also studied the closely-related ANNNI model \cite{milsted_statistical_2015, rahmani_phase_2015-1, rahmani_emergent_2015}, results for which we present in \cite{supplemental}.} to the O'Brien-Fendley (OF) model \cite{obrien_lattice_2018},
\begin{multline} \label{eq:H_OBF}
  H = \sum_{j=1}^N \big[ -\sigma^Z_j \sigma^Z_{j+1} - \sigma^X_j \\
   + \lambda \left(\sigma^X_j \sigma^Z_{j+1} \sigma^Z_{j+2} + \sigma^Z_j \sigma^Z_{j+1} \sigma^X_{j+2} \right)\big],
\end{multline}
which contains the critical Ising model for $\lambda=0$ and, as the latter, is symmetric under $\sigma^Z \rightarrow -\sigma^Z$ and self dual under the Kramers-Wannier duality.
The model remains in the Ising CFT universality class for $0 \le \lambda < \lambda_\TCI$. At  $\lambda_\TCI \approx 0.428$ there is a Tri-Critical Ising (TCI) point \cite{obrien_lattice_2018}, which we confirm by extracting the central charge and some selected scaling dimensions, shown in Table~\ref{tab:extrap}, as we did for the Ising model. This model is particularly interesting for our purposes because, with respect to the Ising CFT, the dominant contribution to the $\lambda$ term comes from the irrelevant $T\overline{T}$ operator \cite{mcgough_moving_2016, cavaglia_t_2016, smirnov_space_2017}. With respect to the TCI CFT, the same term corresponds instead to the relevant primary operator $\varepsilon'$ ($\phi_{1,3}$ in the Kac table \cite{friedan_conformal_1984}), which is known to generate a flow to the Ising CFT 
\cite{zamolodchikov_renormalization_1987, ludwig_perturbative_1987}. This can be confirmed by computing the matrix elements of the $\lambda$ term in the low-energy eigenbasis of $H$ at the Ising and TCI points.
 The interpolating flow between the TCI and the Ising CFTs via closely related operators has been studied in integrable field theory \cite{kastor_rg_1989, zamolodchikov_tricritical_1991, klassen_spectral_1992} and in integrable lattice models \cite{feverati_lattice_2002, pearce_excited_2003, feverati_integrals_2004}, as well as using the truncated CFT Hilbert space approach \cite{lassig_scaling_1991, giokas_renormalisation_2011}. Here, we study the flow nonperturbatively in a nonintegrable lattice model using methods that can be applied to any spin-chain system.
To do this, we compute the low-energy spectrum of the model for fixed $\lambda$, scaled and shifted so that the ground state has $E = \Delta_I = 0$ and $|T\rangle$ has $E = \Delta_T = 2$ (see \cite{milsted_extraction_2017}), as a function of the system size~$N$.

We call the flow with $N$ a \emph{spectral} RG flow \footnote{Notice that the ratio $1/N$ of the lattice spacing to the system size can be understood as a UV length scale, hence the flow with $N$ can be considered an RG flow.} to emphasize that we are studying the flow of the low-energy spectrum, rather than the couplings of an effective Hamiltonian. How should we expect the spectral RG flow to look? We can think of the model with $\lambda = \lambda_\TCI  - \delta$ (for small $\delta>0$) as a relevant deformation of the TCI CFT. Accordingly, at small $N$ the low-energy physics will be dominated by the nearby TCI point, while increasing $N$ will eventually reveal the Ising CFT. We observe this flow at e.g.\ $\lambda=0.4$, where in Fig.~\ref{fig:4specs} we see that the low-energy excitations spectrum at $N=32$ exhibits some striking similarities to the TCI CFT spectrum, while at $N=256$ it looks like the Ising CFT spectrum. Also in Fig.~\ref{fig:4specs}, we show conformal tower membership computed using~\eqref{Hn} \cite{milsted_extraction_2017, supplemental}. At $N=32$, despite strong corrections due to the relevant $\varepsilon'$ perturbation and further irrelevant perturbations, we nevertheless reproduce the low-lying tower-membership results of the TCI. At large $N$, the state identifications match the Ising CFT.

\begin{figure}
  \includegraphics[width=\linewidth]{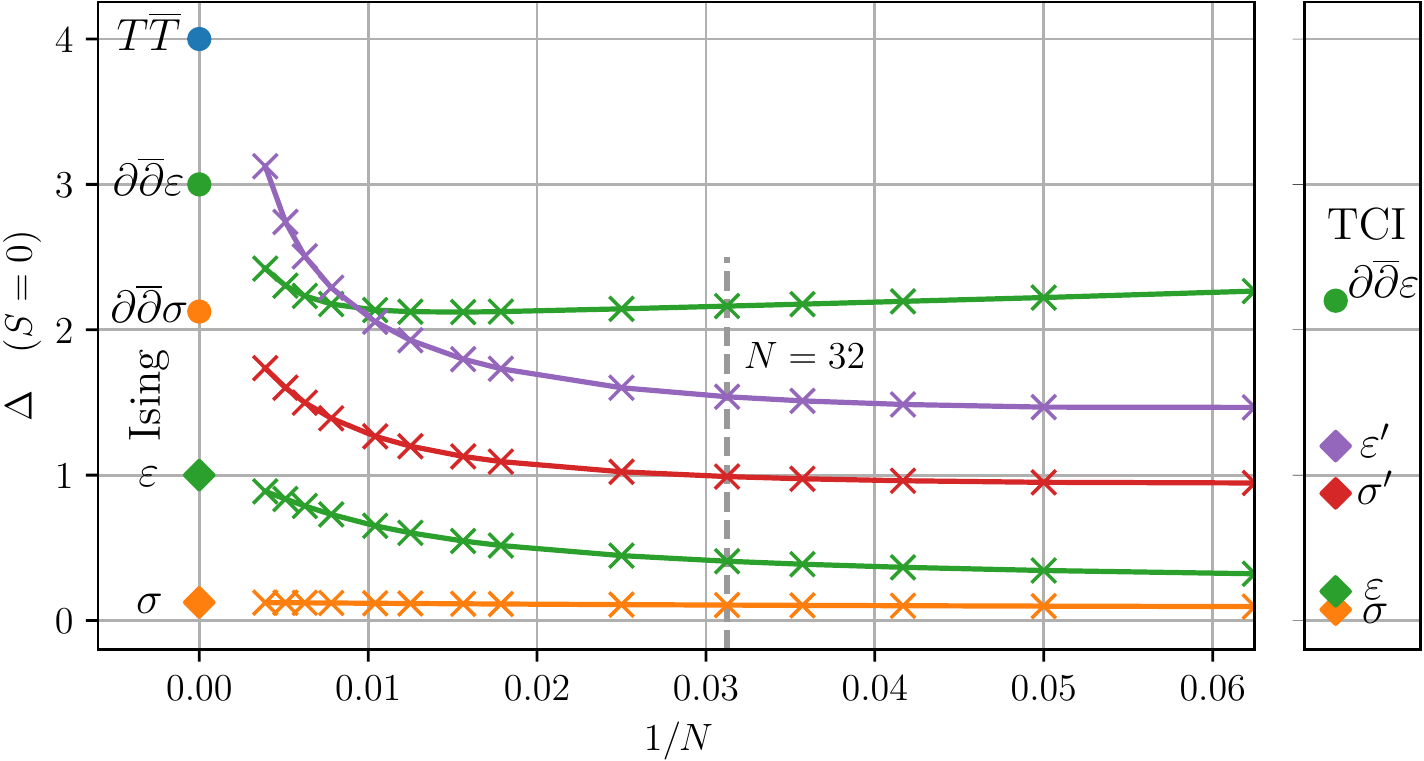}\\
  \vspace{-0.8em}
  \caption{\label{fig:flow_N} Spectral RG flow (crosses) of the first 5 energy levels (as apparent scaling dimensions $\Delta$) at momentum zero, excluding $\Delta=0$, extracted from the OF model with $\lambda=0.4$, using puMPS with $D \le 52$. For comparison, we also plot the exact scaling dimensions of the Ising and TCI CFTs (dots, diamonds). The crossover between the two highest levels plotted, which we confirm by tracking conformal tower membership using $H_n$ matrix elements, is consistent with these states belonging to different Kramers-Wannier self-duality sectors.}
\end{figure}

In Fig.~\ref{fig:flow_N} we further plot the spectral RG flow at $\lambda=0.4$ for a selection of states, including some that would correspond to primary operators in the TCI CFT. We find we can easily determine which Ising CFT operators the TCI CFT primaries are mapped to:
\begin{center}
\begin{tabular}{r|c|c|c|c|c}
TCI operator & $\:I\:$ & $\:\sigma\:$ & $\:\varepsilon\:$ & $\:\sigma'\:$ & $\:\varepsilon'\:$ \\
\hline
Ising operator & $I$ & $\sigma$ & $\varepsilon$ & $\partial \overline{\partial}\sigma$ & $T\overline{T}$ \\
\end{tabular}
\end{center}
These results match those found in other studies of different microscopic realizations of the same CFTs, e.g.~\cite{pearce_excited_2003}, and conform with expectations from symmetry considerations. The identity of $\varepsilon'$ in the TCI CFT with $T\overline{T}$ in the Ising CFT matches their both being associated with the $\lambda$ term in $H$.

\begin{figure}
  \includegraphics[width=\linewidth]{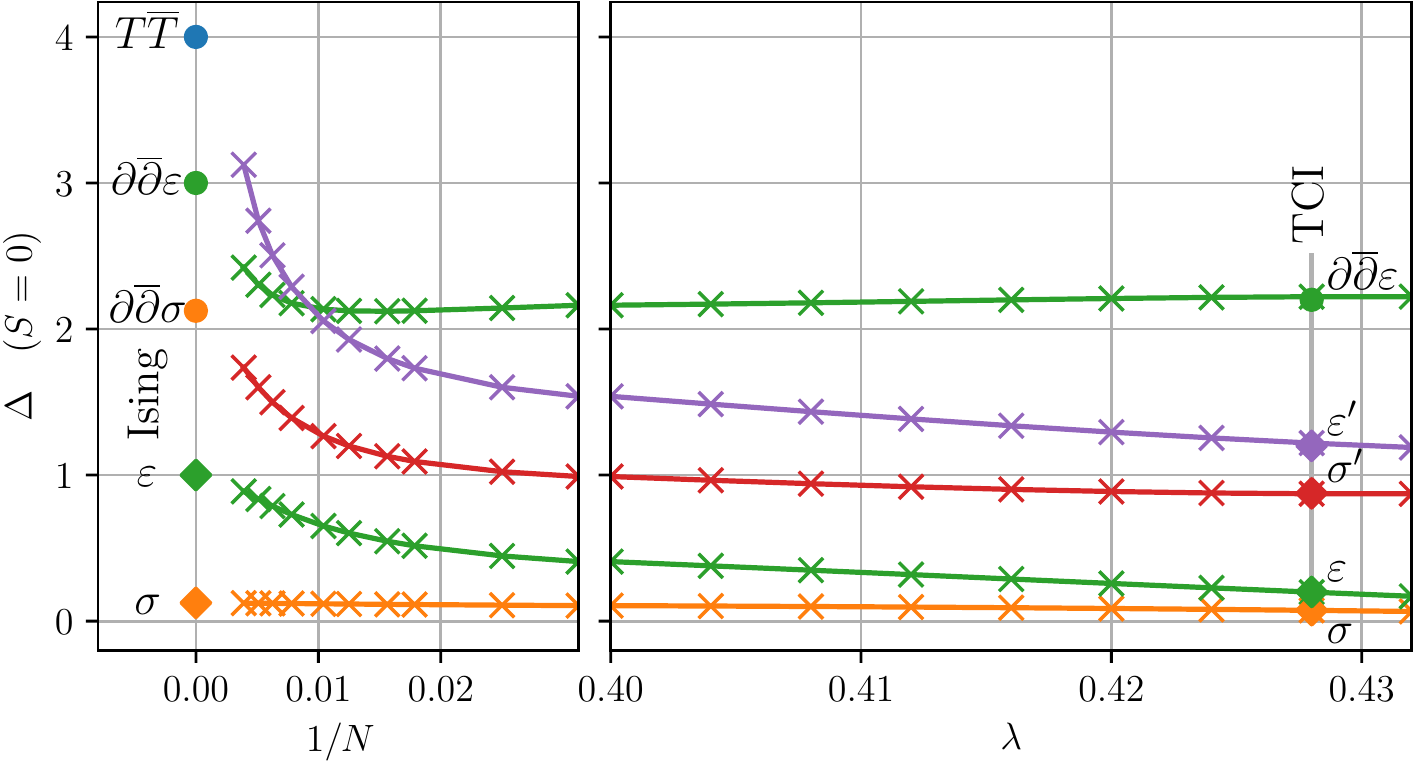}\\
  \vspace{-0.8em}
  \caption{\label{fig:flow_ham} Connection of the spectral RG flow of Fig.~\ref{fig:flow_N} (left) to the ``flow'' of OF model energy levels as a function of $\lambda$ at fixed system size $N=32$, computed using puMPS with $D=28$. Note how the apparent scaling dimensions agree with the TCI CFT values at the TCI point $\lambda_\TCI \approx 0.428$.}
\end{figure}

We can better confirm the TCI operator identities of the low-energy states at $\lambda = 0.4$ by tracking them as a function of $\lambda \rightarrow \lambda_\TCI$. This we do in Fig.~\ref{fig:flow_ham} for fixed $N=32$. We find a very similar pattern to Fig.~\ref{fig:flow_N}, which we would expect if the RG flow of Hamiltonian couplings sends $\lambda$ to zero for any starting $\lambda < \lambda_\TCI$. Using both plots we can connect the low-energy eigenstates at $\lambda=0.4, N=256$, which we identified with Ising CFT operators, with corresponding eigenstates at $\lambda_{\TCI}, N=32$, where they clearly match up with TCI CFT operators.

\begin{figure}
  \includegraphics[width=\linewidth]{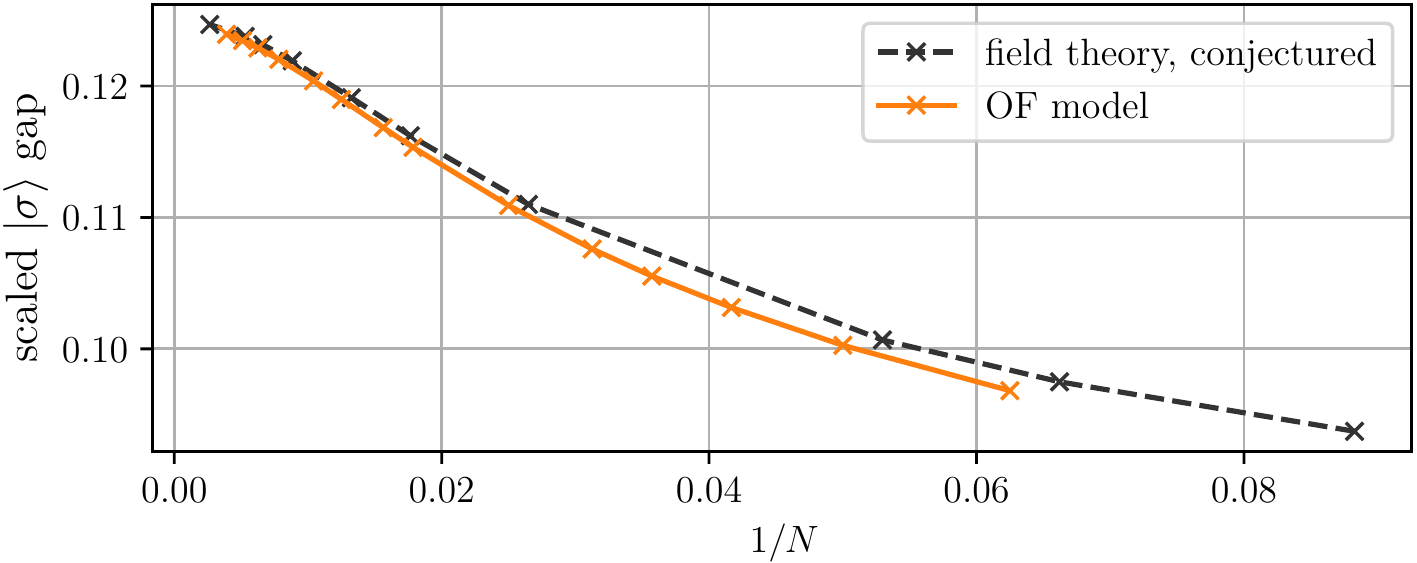}\\
  \vspace{-0.8em}
  \caption{\label{fig:flow_N_vs_ana} Flow of the first spectral gap from Fig.~\ref{fig:flow_N} compared \cite{supplemental} with the integrable field theory result of \cite{klassen_spectral_1992}, conjectured to describe the equivalent flow in the continuum.}
\end{figure}

Finally, in Fig.~\ref{fig:flow_N_vs_ana} we compare \cite{supplemental} our spectral RG flow to the results of \cite{klassen_spectral_1992}, where methods of integrable field theory are used to arrive at a conjecture for the RG flow of the first spectral gap in the continuum. We find increasingly good agreement for larger system sizes $N \rightarrow \infty$, consistent with vanishing finite-size corrections due to lattice effects. We note that our methods should allow us to study nonperturbatively the RG flow of a large number of additional energy levels in \emph{generic} spin chain systems.

\emph{Summary and conclusions.}
We have proposed and demonstrated the use of puMPS and puMPS Bloch states for extraction of conformal data from critical spin chains. The ability to compute accurate variational low-energy eigenstates at large system sizes (far beyond the reach of exact diagonalization) using these techniques enabled us to study a spectral RG flow in the O'Brien-Fendley model \cite{obrien_lattice_2018} and identify low-energy eigenstates with CFT operators in both the Ising and Tri-Critical Ising CFTs.

We remark that it is \emph{a priori} far from obvious that puMPS Bloch states should be an appropriate ansatz for all low-energy excited states. After all, in a noncritical spin chain only single-particle excitations are well captured by this type of ansatz \cite{haegeman_2013_el_ex}, and a different ansatz \cite{vanderstraeten_2015} is needed to capture multiparticle excitations. 
However, in a critical system (for sufficiently large bond dimension \cite{pirvu_matrix_2012-1}) correlations in the puMPS are long range so that the tensor $B$ of \eqref{puMPS_ex} is capable of modifying the ground state wavefunction \emph{more globally} than in the gapped case, making the ansatz more expressive. Note that the ansatz can easily be further improved by considering $B$ tensors that encompass two or more lattice sites, instead of one \cite{haegeman_2013_post}.

Finally, we comment on the benefits of dealing with variational energy eigenstates, such as puMPS Bloch states, that are exact momentum eigenstates by construction. Perhaps most importantly, the momentum directly delivers the conformal spin via \eqref{circle_spec}, which is therefore known exactly. Furthermore, distinguishing between degenerate energy eigenstates via momentum makes it easier to isolate states corresponding to particular CFT operators. This is crucial for follow-up work \cite{zou_upcoming_2018} in which we variationally determine lattice operators corresponding to CFT primary field operators, allowing us to compute operator product expansion (OPE) coefficients for primary fields, thus completing the extraction of conformal data from a generic critical quantum spin chain Hamiltonian.

\begin{acknowledgments}
  \emph{Acknowlegments.} We thank Martin Ganahl for many useful discussions, as well as Jutho Haegeman and Frank Verstraete for valuable comments. We also thank one of referees for proposing the O'Brien-Fendley model. GV thanks the Insitut des Hautes Études Scientifiques (IHES) for hospitality during the workshop ``Hamiltonian methods in strongly coupled Quantum Field Theory''. The authors acknowledge support from the Simons Foundation (Many Electron Collaboration) and Compute Canada. Research at Perimeter Institute is supported by the Government of Canada through Industry Canada and by the Province of Ontario through the Ministry of Research and Innovation.
\end{acknowledgments}

\bibliography{puMPS_CFT}

\clearpage

\appendix
\section{Supplemental Material}

\subsection{Gradient descent for matrix product states on an infinite line}
We optimize a periodic uniform matrix product state (puMPS) on a finite circle by using a gradient descent method, which is analogous to the gradient optimization of MPS with open boundary conditions on an infinite chain. This gradient optimization method was originally proposed in \cite{ganahl_continuous_2016} as an energy minimization algorithm for continuous matrix product states. Here we start by reviewing its analog for MPS on an infinite lattice. Recall that a translation invariant infinite MPS is defined as
\begin{equation}
|\Psi(A)\rangle=\sum_{\vec{s}}v^{\dagger}_L (\cdots A^{s_{-1}}A^{s_0}A^{s_1}\cdots) v_R |\vec{s}\rangle,
\end{equation}
where $s_i=1,2,\cdots d$ labels a basis of the local Hilbert space on the site with position $i$, $\vec{s}=\cdots s_{-1}s_0s_1\cdots$, and $A^{s_i}$  is a set of $d$ matrices with size $D\times D$ that specifies the infinite MPS. $D$ is referred to as the \textit{bond dimension} of the MPS.

The above variational ansatz naturally has a gauge freedom, i.e. two sets of matrices $A^s$ and $A'^s$ describe the same state, i.e. $|\Psi(A)\rangle=|\Psi(A')\rangle$ if they are related by a gauge transformation, $A'^s=g^{-1}A^s g$, where $g$ is a $D\times D$ invertible matrix. We can enforce certain conditions for the tensor $A$ for any translation invariant infinite MPS by exploiting this gauge freedom. One convenient condition is the \textit{left canonical gauge},  where we fix $A=A_L$ that satisfies
\begin{equation}
\label{canonical1}
\sum_{s}A^{\dagger s}_L A^s_L=\mathbf{1}
\end{equation}
and
\begin{equation}
\label{canonical2}
\sum_{s}A^s_L \lambda^2 A^{\dagger s}_L=\lambda^2,
\end{equation}
where $\lambda$ is a $D\times D$ diagonal matrix consisting of descending positive numbers $\lambda_i$ as its diagonal elements. The $\lambda_i$'s are the \textit{Schmidt coefficients} of the bipartition of the infinite chain into left and right semi-infinite chains. Given the original tensor $A$ of the MPS, the left canonical tensor $A_L$ can be obtained with a standard procedure \cite{schollwock_2011}. We will use the left canonical gauge throughout the paper.

Consider a local deformation of the MPS that changes the tensor only on the site $1$ into another tensor. The deformed state 
\begin{equation}
\label{CentralMPS}
|\Psi_{A_L}(A_C)\rangle=\sum_{\vec{s}}v^{\dagger}_L (\cdots A^{s_{0}}_L(A^{s_1}_C\lambda^{-1})A^{s_2}_L \cdots) v_R |\vec{s}\rangle
\end{equation}
depends on a $d\times D\times D$ tensor $A_C$, which is referred to as the \textit{central tensor}. The choice of $A_C$ to parametrize the local deformation is justified by \eqref{effnorm_obc} below. When $A_C=A_L \lambda$, the state $|\Psi_{A_L}(A_C)\rangle$ comes back to the original state $|\Psi(A_L)\rangle$.

Recall that the expectation value of any one site operator $O$ on site $1$ for the deformed state is $\langle O(1) \rangle = \mathrm{Tr}(O_{ss'}A^s_C A^{s'\dagger}_C)$, where repeated upper and lower indices are implicitly summed. In particular, the square of the norm of the deformed MPS equals the vector norm of tensor $A_C$,
\begin{equation}
\label{effnorm_obc} 
\langle \Psi_{\bar{A}_L}(\bar{A}_C)|\Psi_{A_L}(A_C)\rangle=\delta_{\mu\nu} \bar{A}^{\mu}_CA^{\nu}_C,
\end{equation}
where we use $\mu=(s,a,b)$ to denote the combination of the physical index $s$ and matrix indices $(a, b)$, that is $A^\mu\equiv (A^s)_{ab}$. This leads to a crucial simplification of the algorithm for gradient optimization of MPS with open boundary conditions. We will see in the next section that the norm of a locally deformed puMPS is related to the central tensor $A_C$ not by an identity matrix, but a general positive definite matrix, which results in additional complexity in the gradient algorithm for puMPS.

 Now we explain how gradient optimization works. To optimize the MPS for the ground state, we minimize the energy function
\begin{equation}
\label{E}
E(A_L,\bar{A}_L)=\frac{\langle\Psi(\bar{A}_L)|H|\Psi(A_L)\rangle}{\langle\Psi(\bar{A}_L)|\Psi(A_L)\rangle}
\end{equation}
with respect to the left canonical tensor $A_L$. However, we will not minimize this highly nonlinear functional directly. Instead, we will work with an auxiliary energy function
\begin{equation}
\label{auxE}
E_{A_L}(A_C,\bar{A}_C)=\frac{\langle\Psi_{\bar{A}_L}(\bar{A}_C)|H|\Psi_{A_L}(A_C)\rangle}{\langle\Psi_{\bar{A}_L}(\bar{A}_C)|\Psi_{A_L}(A_C)\rangle}
\end{equation}
which only depends on the central tensor $A_C$ on site 1. The auxiliary energy function satisfies following properties: (i) $E_{A_L}(A_C,\bar{A}_C)=E(A_L,\bar{A}_L)$ if $A_C=A_L \lambda$. (ii) Under an infinitesimal change of tensor $A'_L=A_L+\delta A_L$, the change in the original energy function $\delta E(A_L,\bar{A}_L) \equiv E(A'_L,\bar{A}'_L)-E(A_L,\bar{A}_L)$ is related to the change in the auxiliary energy function $\delta E_{A_L}(A_C,\bar{A}_C) \equiv E_{A_L}(A'_C,\bar{A}'_C)-E_{A_L}(A_C,\bar{A}_C)$ by
\begin{equation}
\label{EErelation}
\delta E(A_L,\bar{A}_L)=N\delta E_{A_L}(A_C,\bar{A}_C)+O((\delta A_L)^2)
\end{equation}
if $A_C=A_L \lambda$ and $A'_C=A'_L \lambda$, where $N$ is the size of the system. Thus, if an infinitesimal change $\delta A_C$ away from $A_C=A_L \lambda$ decreases the auxiliary energy function, \eqref{auxE}, the corresponding change $\delta A_L=\delta A_C \lambda^{-1}$ also decreases the original energy function, \eqref{E}. We expect that this still works for a finite but small change. In practice, we find the direction $\Delta A_L$ in which $A_L$ changes that decreases \eqref{E}, and set the step size by a line search or based on empirical observations.

\begin{figure}
  \includegraphics[width=7cm]{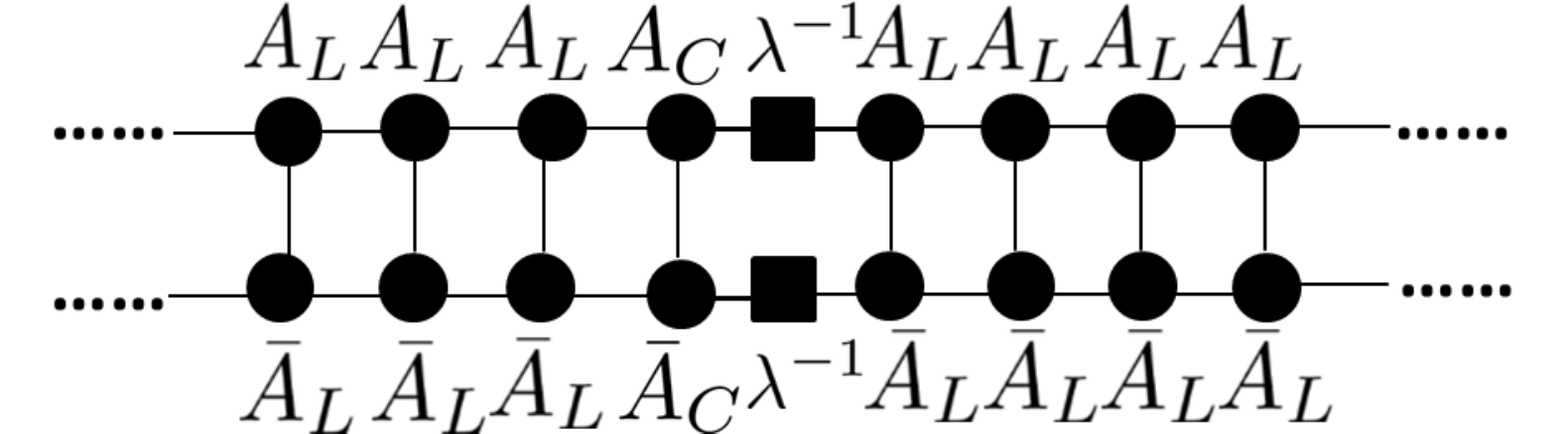}
  \vspace{0.15cm}\\
  \includegraphics[width=7cm]{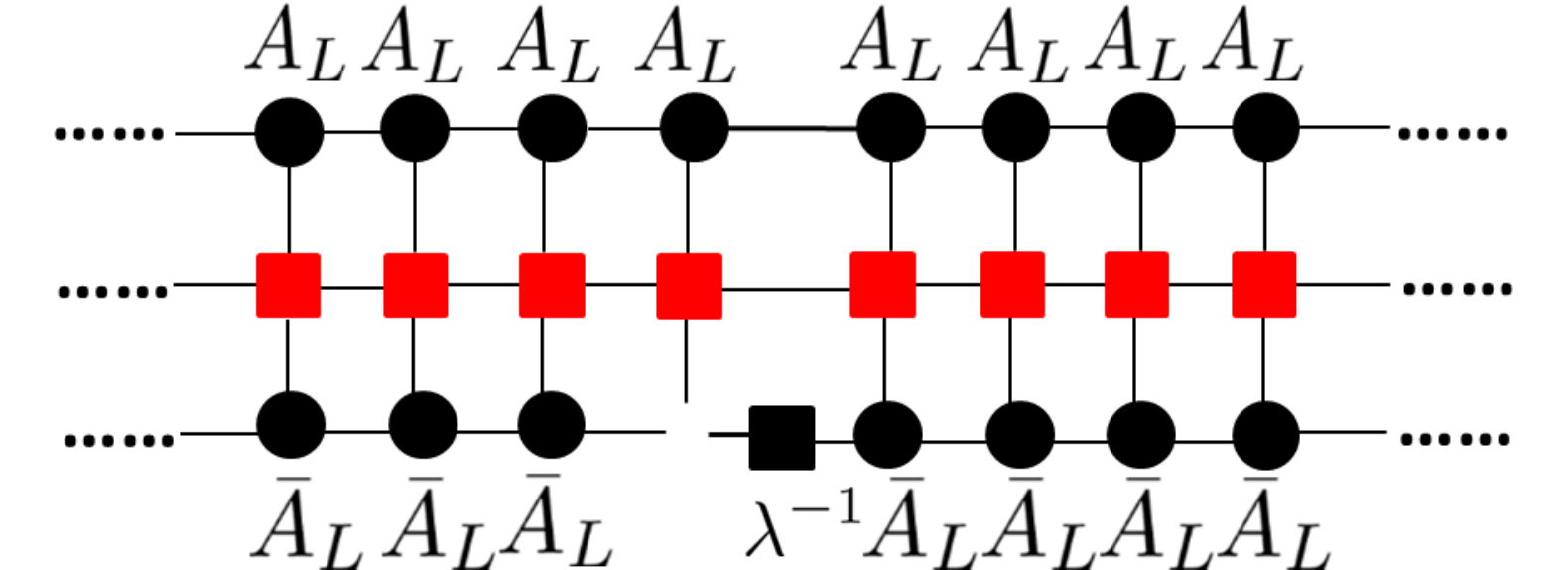}
\caption{\label{fig:local_gr_obc} The tensor networks for the derivative of the auxiliary energy function with respect to the central tensor in \eqref{energy_derivative} for an infinite translation invariant MPS. Top: the denominator, i.e. the square of the norm of locally deformed puMPS with central tensor $A_C$. It equals the vector norm of the tensor $A_C$. Bottom: the numerator, where red tensors form a matrix product operator representation of the shifted Hamiltonian $\tilde{H}$.}
\end{figure}

Therefore, to minimize the energy function iteratively, we identify the direction of change $\Delta A_C$ in $A_C$ that decreases \eqref{auxE}, and then change $A_L$ accordingly. More specifically, given an initial MPS, we compute its left canonical tensor $A_L$ and the diagonal matrix $\lambda$. Next, taking the derivative of $E_{A_L}(A_C,\bar{A_C})$ with respect to the central tensor $\bar{A}_C$ on site 1, we obtain the local gradient $\partial E_{A_L}(A_C,\bar{A}_C)/\partial \bar{A}^\nu_C$. The computation of the local gradient is easier if we use a shifted Hamiltonian $\tilde{H}=H-\langle\Psi(\bar{A}_L)|H|\Psi(A_L)\rangle$ instead of $H$, which leads to
\begin{equation}
\label{energy_derivative}
\frac{\partial E_{A_L}(A_C,\bar{A}_C)}{\partial \bar{A}^{\nu}_C}=\frac{\langle (\partial/\partial \bar{A}^{\nu}_C) \Psi_{\bar{A}_L}(\bar{A}_C)|\tilde{H}|\Psi_{A_L}(A_C)\rangle}{\langle\Psi_{\bar{A}_L}(\bar{A}_C)|\Psi_{A_L}(A_C)\rangle}.
\end{equation}
Contracting the tensor network representing \eqref{energy_derivative} gives us the local gradient, see Fig. \ref{fig:local_gr_obc}.
The substitution of $H$ with $\tilde{H}$ is justified by noting that the shift of Hamiltonian does not alter its ground state. The direction $\Delta A_C$ is then chosen as opposite to the direction of the local gradient,
\begin{equation}
\label{optimal_dir_obc}
\Delta A^{\mu}_C= - \delta^{\mu\nu}\frac{\partial E_{A_L}(A_C,\bar{A}_C)}{\partial \bar{A}^{\nu}_C}.
\end{equation}
The appearance of $\delta^{\mu\nu}$ is related to the fact that the vector space of $A_C$ inherits a flat metric from the norm of the locally deformed MPS, as already noted in \eqref{effnorm_obc}. Next, we change $A_L$ in the direction of $\Delta A_C \lambda^{-1}$,
\begin{equation}
A'_L = A_L + \alpha  (\Delta A_C \lambda^{-1})
\end{equation}
where $\alpha>0$ is a step size obtained by either a line search or empirical observations to optimally minimize the energy function. Finally, we replace the tensor $A$ of the MPS with $A'_L$, and use the standard procedure \cite{schollwock_2011} to put it back to the left canonical form. Notice that $A'_L$ does not fulfill the left canonical condition \eqref{canonical1}\eqref{canonical2} in general. In the next iteration the tensor $A_L$ in the auxilary energy function changes to the left canonical tensor of the updated MPS.

We iterate the steps above until the \textit{norm of the gradient}
\begin{equation}
\eta=\sqrt{\delta^{\mu\nu}\frac{\partial E_{A_L}(A_C,\bar{A}_C)}{\partial A^{\mu}_C}\frac{\partial E_{A_L}(A_C,\bar{A}_C)}{\partial \bar{A}^{\nu}_C}}
\end{equation}
is sufficiently small. Notice that $\eta$ would vanish if we had reached the minimum of the energy function. In practice, we observe that the error in ground state energy is on the order of $\eta^2$, thus we may stop when $\eta$ equals the square root of the expected precision in energy.

The most costly part of the algorithm is the computation of the local gradient \eqref{energy_derivative} at $\mathcal{O}(D^3)$ per iteration, comparable to the cost of the infinite density matrix renormalization group (iDMRG) \cite{McCulloch2008IDMRG,schollwock_2011} with the additional advantage of keeping explicit translation invariance. Other optimization schemes, such as the infinite time evolution block decimation (iTEBD) \cite{Vidal2007ITEBD} and the time dependent variational principle (TDVP) \cite{haegeman_2011, pirvu_matrix_2012}, though keep translation invariance explicitly, follow an imaginary time evolution trajectory which converges slower than the gradient optimization.

We finally remark on the extraction of conformal data with infinite MPS. It is well known that when applying infinite MPS to a critical system the finite bond dimension $D$ introduces an artificial finite correlation length $\xi(D)$ which grows with $D$. Despite the fact that long distance physics beyond the correlation length is not captured,  algebraically decaying correlation functions at shorter distances can be faithfully reproduced, and conformal data such as scaling dimensions and the central charge can be accurately extracted \cite{stojevic_2015}.
\subsection{Gradient descent for puMPS}
A puMPS on a finite circle is a finite size analog of the translation invariant MPS on an infinite line. A puMPS on $N$ sites is defined as
\begin{equation}
|\Psi(A)\rangle=\sum_{\vec{s}=1}^d\mathrm{Tr}(A^{s_1}A^{s_2}\cdots A^{s_N})|\vec{s}\rangle,
\end{equation}
where $\vec{s} = s_1\dots s_N$ and $d$ is the dimension of the Hilbert space for a single site. It can represent the ground state of critical spin chains with high fidelity provided that the bond dimension $D$ grows polynomially with the system size $N$ \cite{verstraete_2006}.

The optimization of a periodic MPS, which typically costs $\mathcal{O}(ND^5)$ or higher, is numerically more costly than open boundary MPS \cite{verstraete_dmrg_2004, pippan_efficient_2010} which costs $\mathcal{O}(ND^3)$. In a gapped system, one may reduce the cost of optimizing a puMPS to $\mathcal{O}(ND^3)$ by truncating singular values of the transfer matrix \cite{pirvu_matrix_2012}. However, in a critical system, the truncation will introduce larger errors. Therefore, here we will not follow such a strategy. Instead, we propose a local gradient descent method that resembles the gradient optimization for open boundary MPS introduced in the previous section, but with cost $\mathcal{O}(ND^5)$. It also shares some features with the TDVP method although we do not follow an imaginary time evolution trajectory.

The energy minimization goes as follows. First, we treat the tensor $A$ as if it belonged to an infinite MPS and compute $A_L$ and $\lambda$ that satisfy the left canonical condition \eqref{canonical1}\eqref{canonical2}. Then we define the locally deformed puMPS as
\begin{equation}
|\Psi_{A_L}(A_C)\rangle=\sum_{\vec{s}=1}^d\mathrm{Tr}[(A^{s_1}_C\lambda^{-1}) A^{s_{2}}_L\cdots A^{s_{N-1}}_L]|\vec{s}\rangle.
\end{equation}
Here, as in \eqref{CentralMPS} for the infinite MPS case, the dependence on $A_C$ is only on site $1$.

\begin{figure}
  \begin{equation}
   \nonumber
    g_{\mu\nu} = \vcenter{\hbox{\includegraphics[width=7cm]{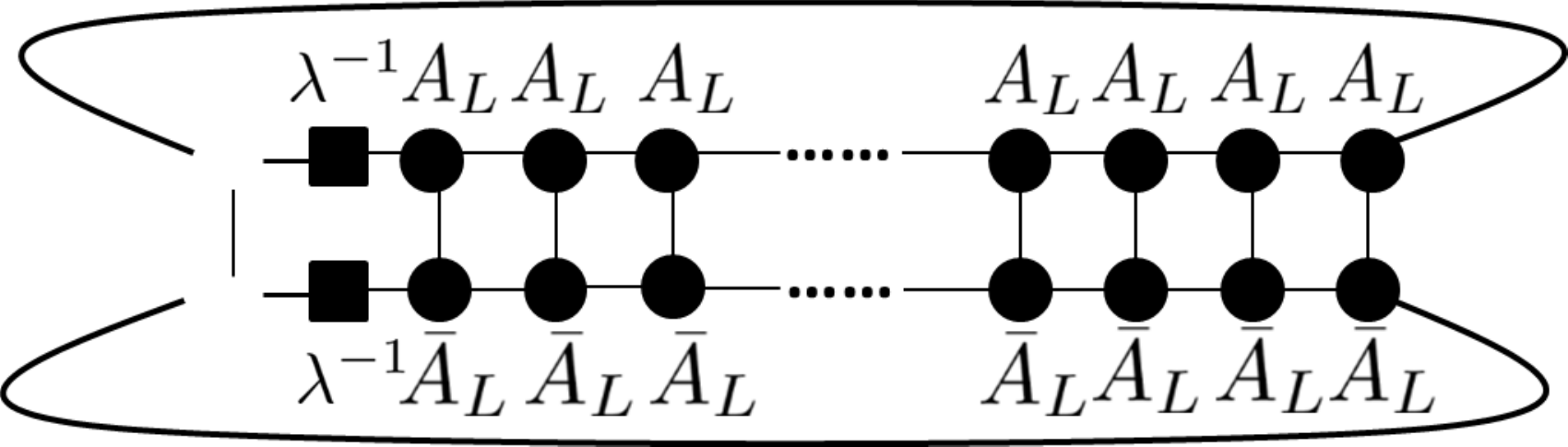}}}
  \end{equation}
  \vspace{0.1cm}\\
\begin{equation}
\nonumber
  \frac{\partial E_{A_L}(A_C,\bar{A}_C)}{\partial \bar{A}^\nu_C}=\vcenter{\hbox{\includegraphics[width=6cm]{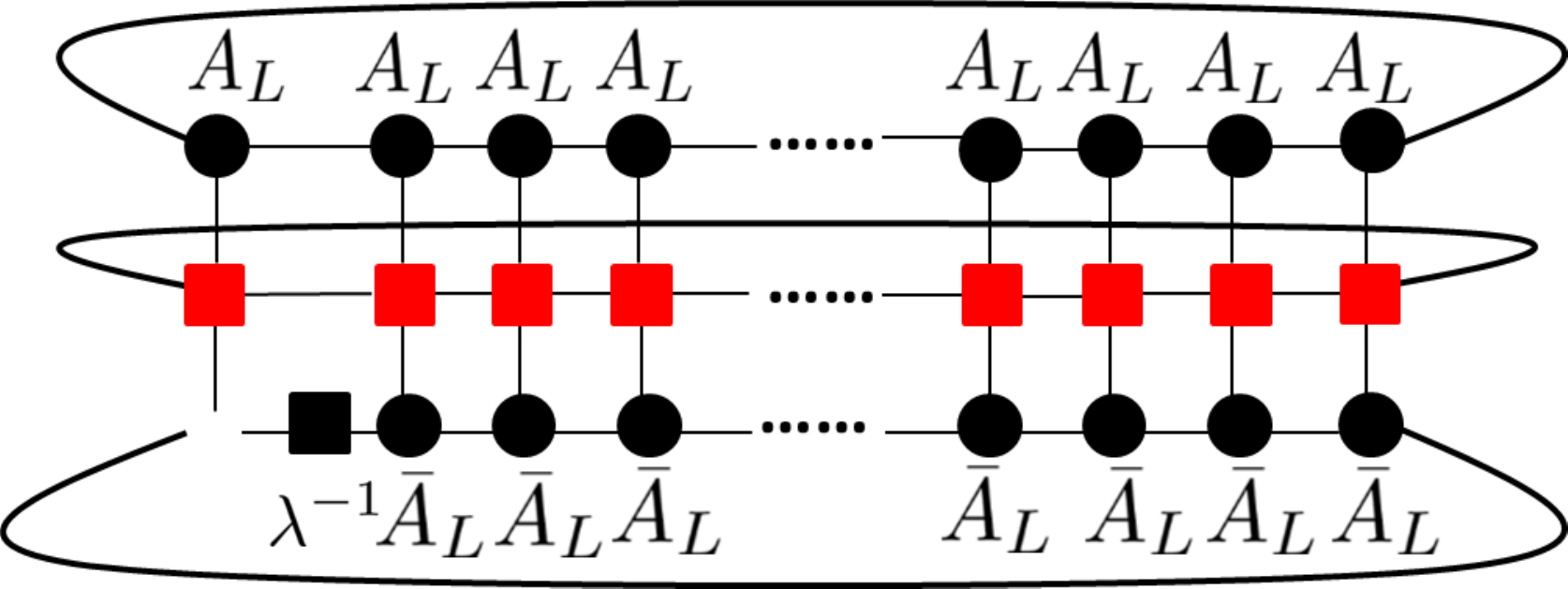}}}
\end{equation}
\caption{\label{fig:local_gr} Top: the tensor network for the local effective norm matrix \eqref{local_eff_norm} for the deformed puMPS in the left canonical gauge. Bottom: the tensor network for the derivative of the auxiliary energy function with respect to the puMPS tensor $A_C$ in \eqref{energy_derivative}, assuming that the puMPS is normalized. The red tensors form a matrix product operator representation of the shifted Hamiltonian $\tilde{H}$.}
\end{figure}

We can relate the square of the norm of the deformed puMPS to the central tensor $A_C$ with a bilinear form,
\begin{equation}
\label{local_eff_norm}
\langle\Psi_{\bar{A}_L}(\bar{A}_C)|\Psi_{A_L}(A_C)\rangle= \bar{A}^{\mu}_C g_{\mu\nu} A^{\nu}_C.
\end{equation}
We call the positive definite matrix $g_{\mu\nu}$ the \textit{local effective norm matrix} for the central tensor. For an infinite MPS, we can read off $g_{\mu\nu}=\delta_{\mu\nu}$ from (\ref{effnorm_obc}). However, $g_{\mu\nu}$ is nontrivial in the case of puMPS as a result of periodic boundary conditions, as represented in Fig.~\ref{fig:local_gr}. Then, as we did for infinite MPS, we can define the auxiliary energy function as \eqref{auxE} with the state substituted by the deformed puMPS. The computation of the local gradient $\partial E_{A_L}(A_C,\bar{A}_C)/\partial \bar{A}^\nu_C$ is also simplified by using the shifted Hamiltonian, which leads to the same expression for the local gradient as (\ref{energy_derivative}) but the state substituted by the deformed puMPS. It is represented as a tensor network in Fig.~\ref{fig:local_gr}. These tensor networks can be contracted with time cost $\mathcal{O}(ND^5)$.

Next, we compute the direction of change $\Delta A_C$ in the central tensor $A_C$ with gradient descent. The optimal direction $\Delta A_C$ that decreases the auxiliary energy function now becomes
\begin{equation}
\label{optimal_gradient}
\Delta A^{\mu}_C = -g^{\mu\nu}\frac{\partial E_{A_L}(A_C,\bar{A}_C)}{\partial \bar{A}^{\nu}_C},
\end{equation}
where $g^{\mu\nu}$ is the inverse of the nontrivial metric $g_{\mu\nu}$ as given by the local effective norm matrix in \eqref{local_eff_norm}, satisfying $g^{\mu\nu}g_{\nu\rho}=\delta^{\mu}_{\rho}$. This results from the fact that the space of $A_C$ now inherits the nontrivial metric $g_{\mu\nu}$ from the norm of the locally deformed puMPS, compared to \eqref{optimal_dir_obc}. The inverse metric, however, does not need to be computed densely, since all we need is to compute \eqref{optimal_gradient}, where the left hand side can be solved for iteratively. The use of the left canonical form, though it does not eliminate the need for inverting the metric, is advantageous in practice, since it generally leads to a better conditioned local metric favored by iterative linear equation solvers.

Finally, we transform the change of $A_C$ into the change of puMPS tensor $A_L$ by $\Delta A_L = \Delta A_C \lambda^{-1}$, then update the tensor according to
\begin{equation}
A'_L = A_L + \alpha \Delta A_L,
\end{equation}
where $\alpha>0$ is the step size obtained by either line search or empirical observations. $A'_L$ is then used as the new puMPS tensor (for all sites), resulting in an updated puMPS. The iteration then restarts from computing the left canonical tensor of the updated puMPS.

Convergence is also monitored using the norm of the local gradient with a modified definition
\begin{equation}
\eta=\sqrt{g^{\mu\nu}\frac{\partial E_{A_L}(A_C,\bar{A}_C)}{\partial A^{\mu}_C}\frac{\partial E_{A_L}(A_C,\bar{A}_C)}{\partial \bar{A}^{\nu}_C}}.
\end{equation} 
We also observe that the error in ground state energy is roughly $\eta^2$ for the puMPS gradient descent algorithm. Thus we stop at $\eta<10^{-6}$ in all the simulations, resulting in a $10^{-12}$ error in the ground state energy which is negligible compared to other sources of errors in conformal data, such as the non-universal subleading finite size corrections.

The above method shares with TDVP the computation of an effective norm matrix and its inverse matrix. The main difference is that while in TDVP we compute the full effective norm matrix to follow the trajectory of an imaginary time evolution, which costs $\mathcal{O}(ND^6)$ per iteration for MPS with periodic boundary conditions, in the local gradient descent method above we compute the local effective norm matrix where the dependence of $A_C$ is only kept explicit on one site. In order to find the ground state, it is not necessary to follow the trajectory of an imaginary time evolution. Instead, in many cases \cite{ganahl_continuous_2016,Stauber2017VUMPS} including our case of puMPS, a simpler local gradient method makes energy minimization faster.

\subsection{Preconditioning of puMPS}
Optimization using gradient descent usually suffers from local minima.  \textit{Preconditioning} is a procedure to find an initial state that approximates the global minimum, with which gradient descent converges faster. In the context of puMPS optimization, we observe that starting with a random state only works well for small bond dimension in small systems. For puMPS with larger bond dimension in larger systems, the energy landscape of the variational manifold becomes more complicated, and the algorithm is more likely to get stuck in a local minimum. Here, we use several simple ways of preconditioning.

First, we can directly use the optimized puMPS tensor for system size $N_0$ as the tensor $A$ for the initial state for slightly larger system sizes $N_1>N_0$ with the same bond dimension.

Second, for the same system size, as an initial puMPS state with bond dimension $D_1$, we can use the optimized puMPS tensor with smaller bond dimension $D_0$, enlarging it to $d\times D_1\times D_1$ and filling the vacancies with small random numbers.

\begin{figure}
  \includegraphics[width=\linewidth]{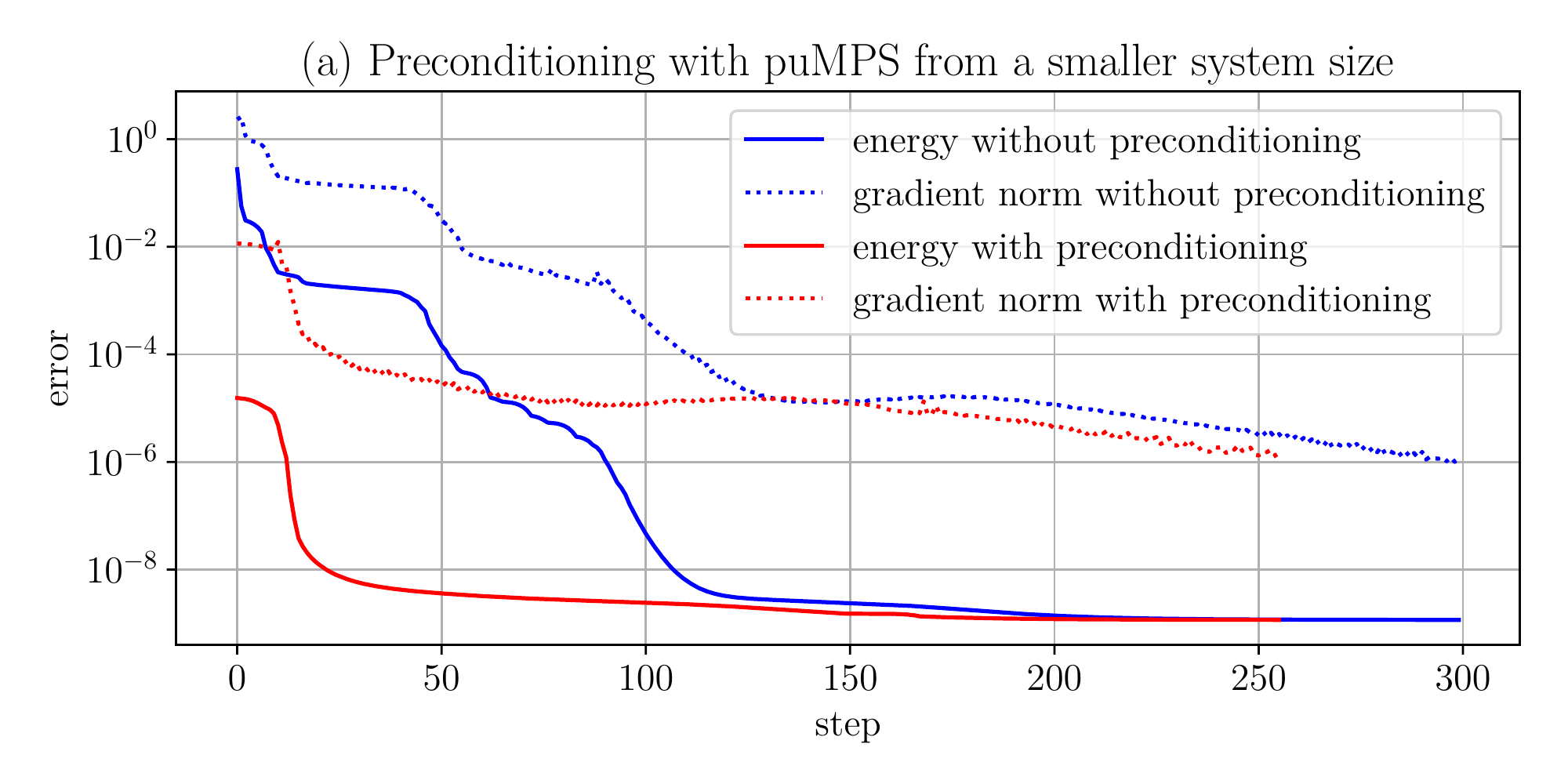}
  \includegraphics[width=\linewidth]{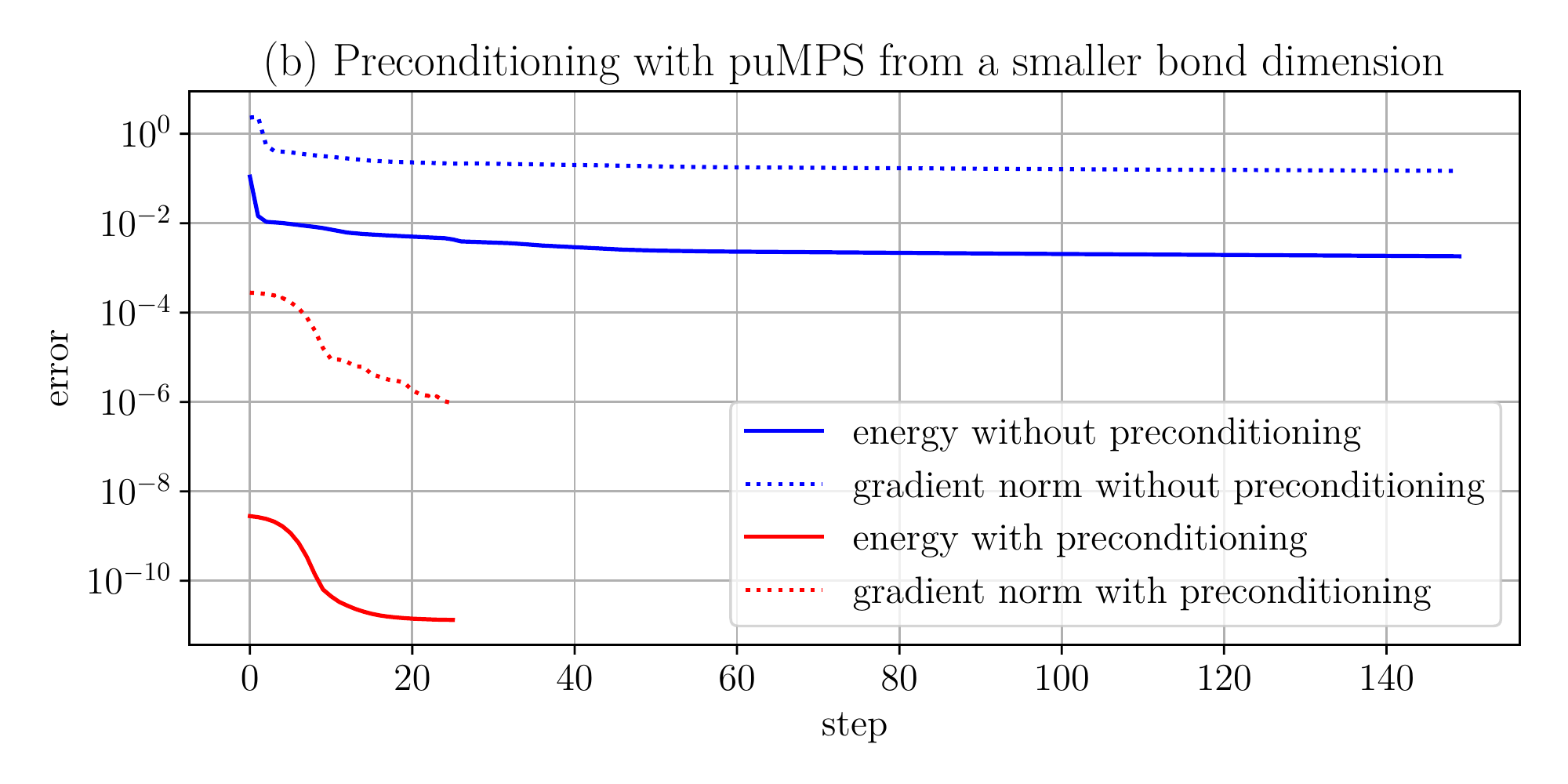}
  \caption{Convergence of the puMPS gradient descent algorithm with preconditioning for the critical Ising model with $N=128$. The dashed lines are the convergence of the gradient norm $\eta$, and the solid lines represent the energy difference of puMPS from the analytic ground state energy at each step of energy minimization. (a) Bond dimension $D=18$, initial state chosen with the pre-optimized puMPS tensor for $N=64,D=18$ (red), and random initial state (blue). (b) Bond dimension $D=30$, initial state chosen by enlarging the optimized puMPS tensor from $D=18,N=128$ (red), and random initial state (blue). Iterations are stopped when $\eta<10^{-6}$. \label{fig:puMPS_conv} }
\end{figure}

We compare the convergence of the local gradient descent algorithm in these two cases, with preconditioning or starting with random state, in Fig.~\ref{fig:puMPS_conv}. The results show that preconditioning significantly accelerates convergence and helps produce accurate ground state approximations within a smaller number of iterations.

\subsection{Excited state ansatz}
The excited state ansatz with momentum $p$ is a Bloch state of optimized puMPS, also known as a tangent vector of the puMPS,
\begin{equation}
\label{puMPS_tangent_vec}
|\Phi_p(B,A)\rangle=\sum_{n=1}^N e^{-i p n} \mathcal{T}^n \sum_{\vec{s}=1}^d \mathrm{Tr} \left( B^{s_1} A^{s_2} \dots A^{s_N} \right) |\vec{s}\,\rangle,
\end{equation}
where $\mathcal{T}$ is the translation operator by one site. The Hamiltonian eigenvalue equation in the subspace of tangent vectors then becomes a generalized eigenvalue equation in the parameter space \cite{pirvu_exploiting_2011},
\begin{equation}
\label{generalized_eigv}
H_{\mu\nu}(p)B^{\nu}=E N_{\mu\nu}(p) B^{\nu},
\end{equation}
where $N_{\mu\nu}(p)$, $H_{\mu\nu}(p)$ are the effective norm matrix and the effective Hamiltonian for tangent vectors in each momentum sector, defined as
\begin{eqnarray}
\label{eff_N}
N_{\mu\nu}(p)=\left\langle \frac{\partial}{\partial \bar{B}^{\mu}}\Phi_p(\bar{B},\bar{A})\right|\left.\frac{\partial}{\partial B^{\nu}}\Phi_p(B,A)\right\rangle \\
\label{eff_H}
H_{\mu\nu}(p)=\left\langle \frac{\partial}{\partial \bar{B}^{\mu}}\Phi_p(\bar{B},\bar{A})\right|H\left|\frac{\partial}{\partial B^{\nu}}\Phi_p(B,A)\right\rangle,
\end{eqnarray}
where the derivative is taken with respect to the tensor on \textit{all} sites in contrast with the local effective norm matrix in ground state optimization.

\begin{figure}
  \includegraphics[width=\linewidth]{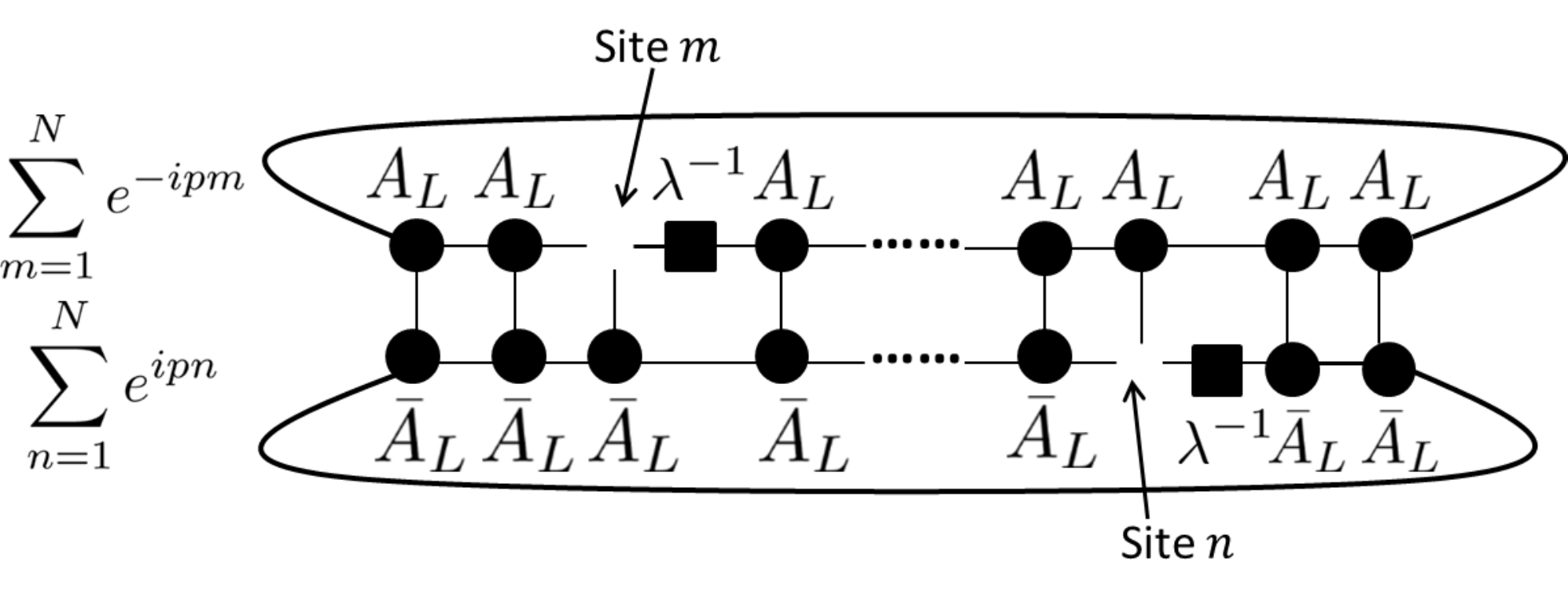}
  \includegraphics[width=\linewidth]{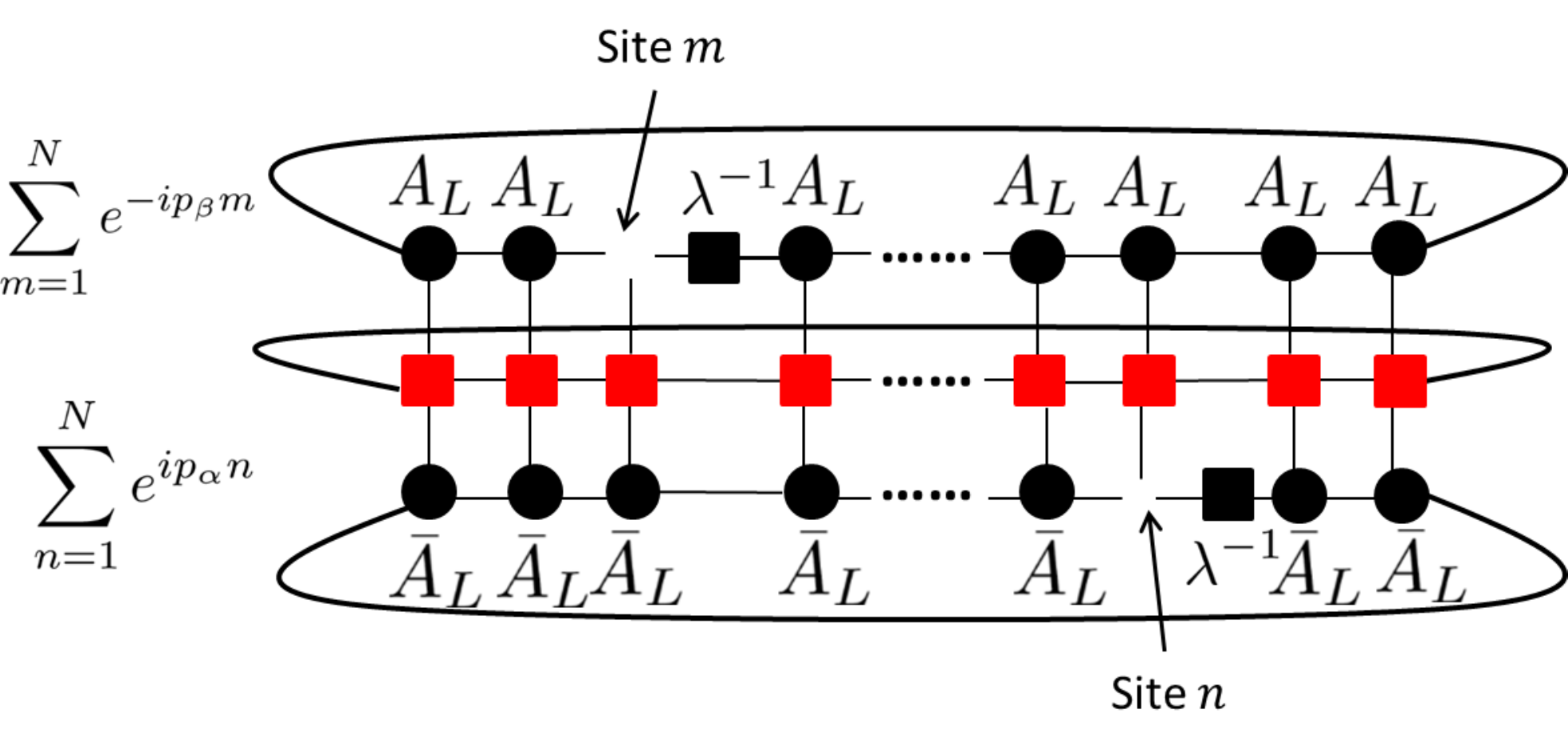}
  \caption{\label{fig:Neff_Hneff} Top: the tensor network for the effective norm matrix $N_{\mu\nu,C}(p)$ for puMPS tangent states parametrized with $A_L$ and $B_C=B\lambda$. Bottom: the tensor network for the effective Hamiltonian $H_{\mu\nu,C}(p)$ when $p_\alpha=p_\beta=p$, or for the effective $H_n$ matrices $H_{n,\mu\nu,C}(p_\alpha,p_\beta)$, where the red tensors in the middle form a matrix product operator representation of the Hamiltonian or its Fourier modes, respectively.}
\end{figure}

We have to be a bit cautious when solving the generalized eigenvalue equation \eqref{generalized_eigv} by multiplying the inverse of $N_{\mu\nu}$ on both sides. This is because (i) the full effective norm matrix $N_{\mu\nu}(p)$ is only semi-positive definite due to gauge freedom of MPS tangent vectors \cite{pirvu_matrix_2012, haegeman_2013_post}, and (ii) it is not well conditioned, even if we project out its null space, as many positive eigenvalues may be close to zero. The first problem is settled if we use the pseudoinverse of the effective norm matrix instead of the ordinary inverse. To solve the second problem, we can again resort to the left canonical form of the puMPS. We parametrize the puMPS with the left canonical tensor $A_L$, and the tangent tensor $B$ is parametrized with $B_C$ by $B=B_C \lambda^{-1}$,
\begin{equation}
\label{puMPSTvec_C}
|\Phi_p(B_C,A_L)\rangle=\sum_{n=1}^N e^{-i p n} \mathcal{T}^n \sum_{\vec{s}=1}^d \mathrm{Tr}( (B^{s_1}_C\lambda^{-1}) A^{s_2}_L \cdots A^{s_N}_L) |\vec{s}\,\rangle.
\end{equation}
Then \eqref{generalized_eigv} can be rewritten in terms of $B_C$,
\begin{equation}
\label{generalized_eigv_cnetral}
H_{\mu\nu,C}(p)B^{\nu}_C=E N_{\mu\nu,C}(p) B^{\nu}_C,
\end{equation}
where $N_{\mu\nu,C}(p)$ and $H_{\mu\nu,C}(p)$ are obtained by substituting the derivatives in \eqref{eff_N}, \eqref{eff_H} by derivatives with respect to $B_C$ and $\bar{B}_C$ and setting $A=A_L$,
\begin{widetext}
\begin{eqnarray}
\label{eff_N_C}
N_{\mu\nu,C}(p)=\left\langle \frac{\partial}{\partial \bar{B}^{\mu}_C}\Phi_p(\bar{B}_C,\bar{A}_L)\right|\left.\frac{\partial}{\partial B^{\nu}_C}\Phi_p(B_C,A_L)\right\rangle \\
\label{eff_H_C}
H_{\mu\nu,C}(p)=\left\langle \frac{\partial}{\partial \bar{B}^{\mu}_C}\Phi_p(\bar{B}_C,\bar{A}_L)\right|H\left|\frac{\partial}{\partial B^{\nu}_C}\Phi_p(B_C,A_L)\right\rangle.
\end{eqnarray}
They are depicted as tensor networks in Fig.~\ref{fig:Neff_Hneff}.
\end{widetext}

\begin{figure}
  \includegraphics[width=\linewidth]{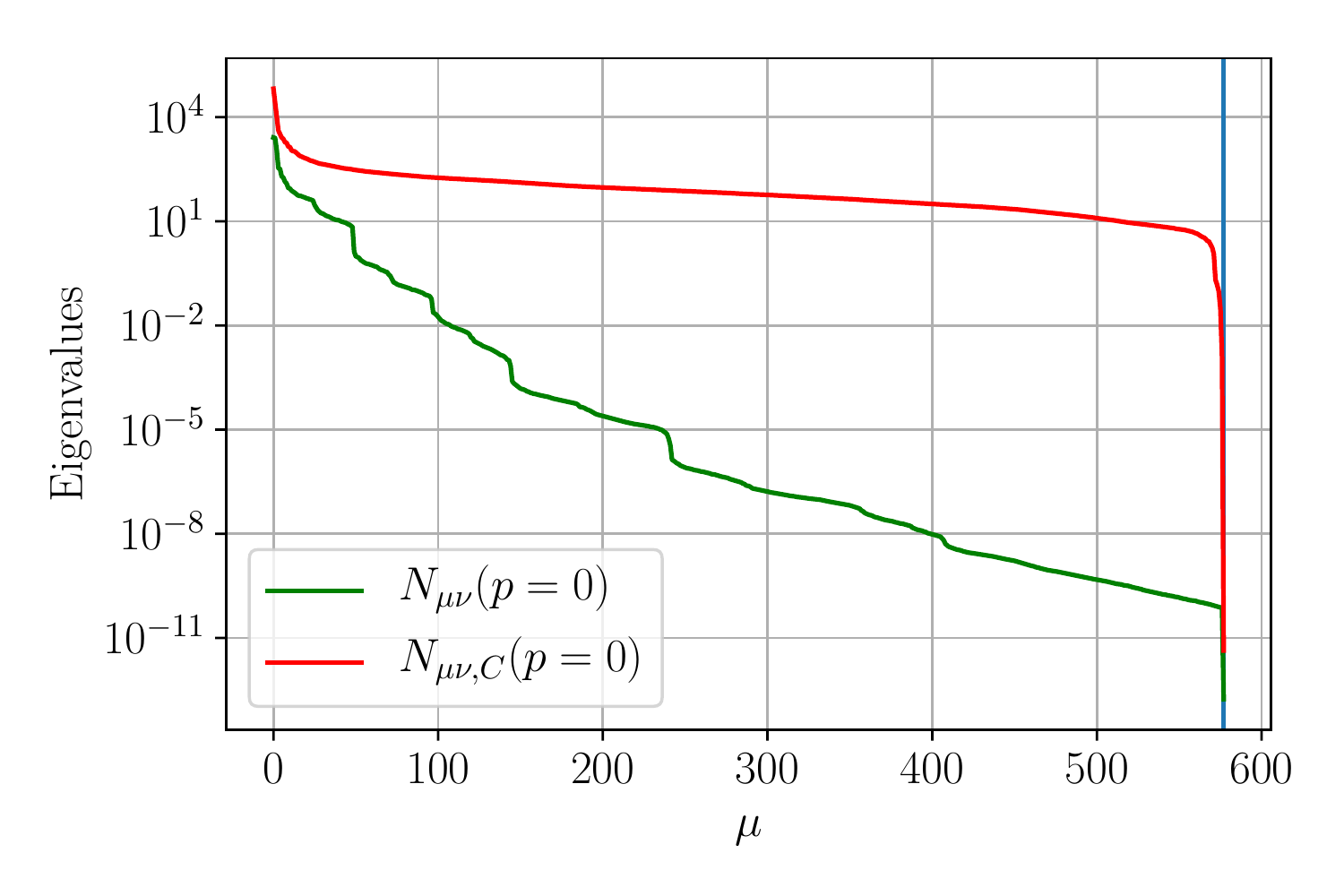}
  \caption{\label{fig:Neff_eig} Nonzero eigenvalues of the effective norm matrix in momentum zero sector of the Ising model with $N=64,D=24$, sorted in descending order. Green: $N_{\mu\nu}(p=0)$ in \eqref{eff_N} where the puMPS tensor $A$ is fixed as the left canonical tensor $A_L$. Red: $N_{\mu\nu,C}(p=0)$ in \eqref{eff_N_C}. The blue vertical line is at $\mu=(d-1)D^2+1$, the number of nonzero eigenvalues resulting from the gauge freedom of puMPS tangent vectors in momentum zero sector.}
\end{figure}

 The effective norm matrix $N_{\mu\nu,C}(p)$ with respect with $B_C$ is much better conditioned than the original effective norm matrix $N_{\mu\nu}(p)$ with respect to $B$ in a random gauge. As an example, we fix $A=A_L$ and plot the eigenvalues of $N_{\mu\nu,C}(p=0)$ and $N_{\mu\nu}(p=0)$ for the Ising model with $N=64$ and puMPS bond dimension $D=24$ to show this explicitly in Fig.~\ref{fig:Neff_eig}.

We then multiply by the \textit{pseudoinverse} $\tilde{N}^{\rho\mu}_C(p)$ of the effective norm matrix on both sides of \eqref{generalized_eigv_cnetral} to obtain the ordinary eigenvalue equation,
\begin{equation}
\label{excited_state_eigv}
\tilde{N}^{\rho\mu}_C(p)H_{\mu\nu,C}(p)B^{\nu}_C=EB^{\rho}_C.
\end{equation}
Finally we compute a set of low-energy eigenvectors in each momentum sector with \eqref{excited_state_eigv} using the Lanczos algorithm, and multiply $B_C$ by $\lambda^{-1}$ to get back to $B$. 

The computation of $N_{\mu\nu,C}(p)$ and $H_{\mu\nu,C}(p)$ costs $\mathcal{O}(ND^6)$. However, since we only need to construct the matrices once for each momentum sector, the actual time cost is usually less than ground state optimization. 
\subsection{Fidelity with Exact Diagonalization}
\begin{figure}
  \includegraphics[width=\linewidth]{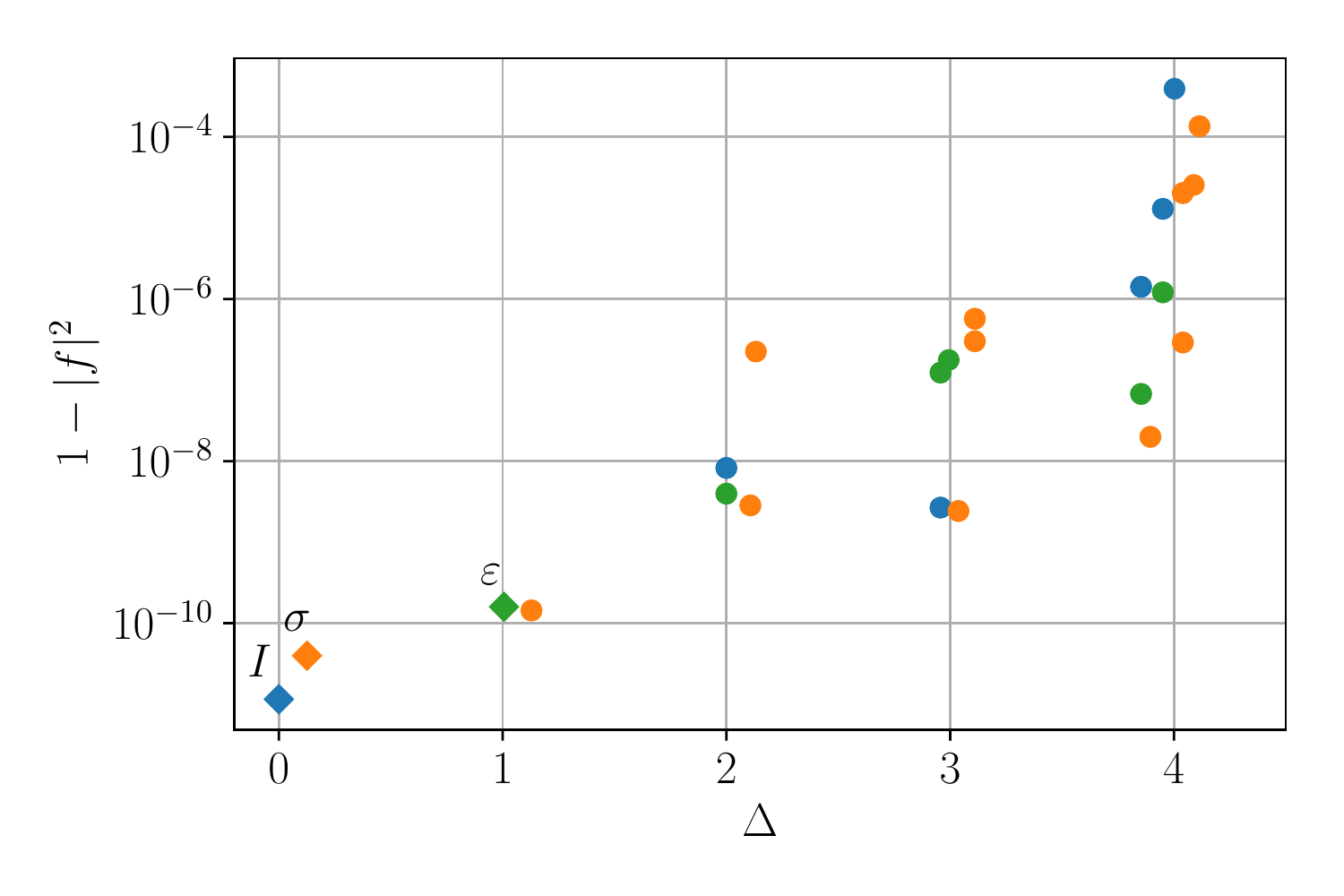}
  \includegraphics[width=\linewidth]{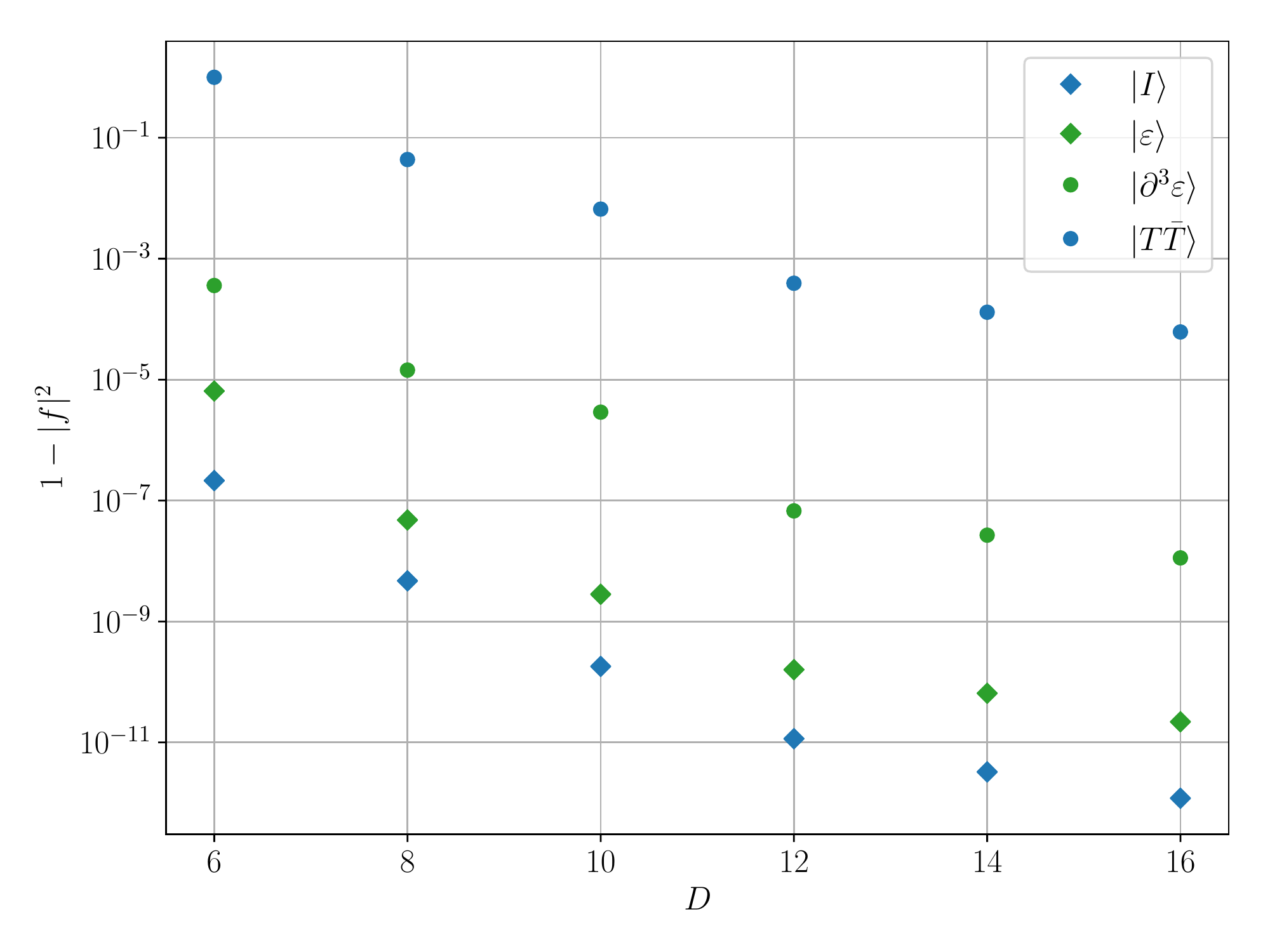}
  \caption{\label{fig:fidelity} Fidelity of the first 41 exactly diagonalized eigenstates of the Ising model ($N=20$) with their variational, puMPS Bloch-state counterparts. Top: fidelity of all 41 states for fixed bond dimension $D=12$. Primary states are labeled with diamonds and descendant states are labeled with dots. Different colors label states in different conformal towers. Bottom: fidelity of four selected states for bond dimensions $6\leq D \leq 16$. All ground states are converged to $\eta<10^{-6}$. }
\end{figure}

In order to check how well the excited states are captured by the above puMPS Bloch state ansatz, we explicitly compute the fidelity of puMPS tangent vectors obtained above with eigenstates obtained by exact diagonaliztion for the Ising model with $N=20$. The fidelity is defined as
\begin{equation}
f_\alpha=\langle \phi_{\alpha}|\phi_{\alpha}^{\mbox{\tiny exact}}\rangle,
\end{equation}
where $|\phi_{\alpha}^{\mbox{\tiny exact}}\rangle$ is the $\alpha$th eigenstate from exactly diagonalizing the full Hamiltonian, and $|\phi_{\alpha}\rangle = |\Phi_{p_\alpha}(B_\alpha,A)\rangle$ is the corresponding eigenstate represented approximately as a puMPS tangent vector. We compute the first 41 eigenstates, corresponding to scaling dimensions $\Delta\leq 4+1/8$. The result is shown in Fig.~\ref{fig:fidelity}.

We can see that although the fidelity decreases as energy increases for a fixed bond dimension, fidelity increases uniformly for each state as the bond dimension increases, regardless of energy and conformal tower of the state. Thus we conclude that the puMPS Bloch state ansatz, \eqref{puMPS_tangent_vec}, can capture all eigenstates in the low-energy subspace with sufficiently small errors, given large enough bond dimension.

\begin{figure}
  \includegraphics[width=\linewidth]{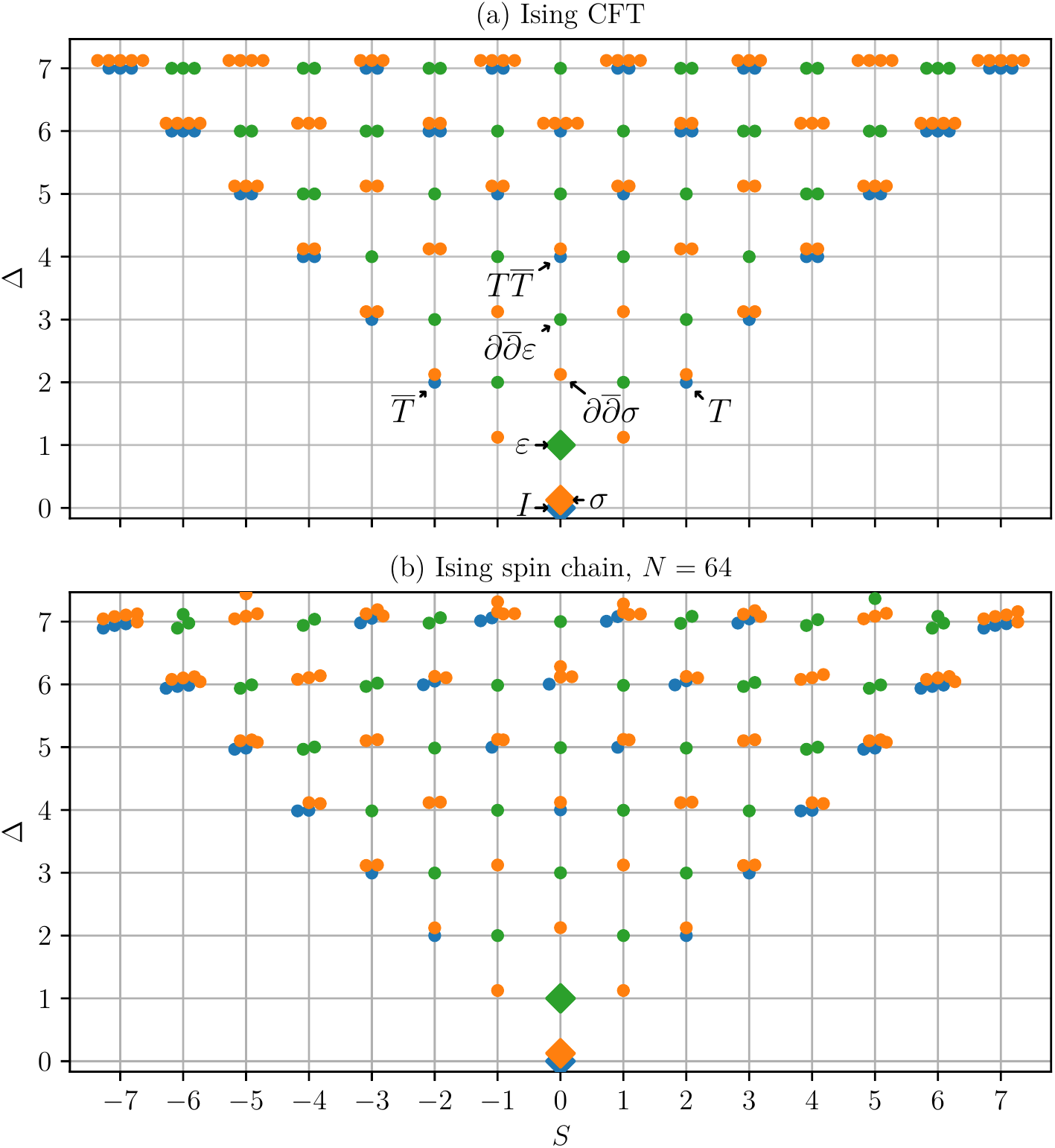}
  \caption{\label{fig:ising_spec} (a) Exact Ising CFT scaling operator spectrum, with diamonds marking primary operators. (b) Ising model spectrum, colored according to numerical conformal-tower identification, for $N=64$ sites using Bloch states on top of a puMPS variational ground state with $D=24$, converged to $\eta<10^{-6}$ (error on ground-state energy density $\approx 10^{-11}$). The numerically-assigned conformal towers are consistent with the CFT result up to level 7, where some states are misidentified. {\bf Note:}\ We displace data points slightly along the x-axis to show degeneracies.}
\end{figure}

\subsection{puMPS Bloch states for generic critical quantum spin chains}

The puMPS Bloch-state ansatz was observed in \cite{pirvu_exploiting_2011} to accurately approximate excited states in two integrable models. In particular, they showed that excitations energies are accurately reproduced in the critical Ising model and the spin-1/2 Heisenberg antiferromagnet and that, in the latter model, a selection of the variational excited states have good fidelity with their exact counterparts.

Here we have extended these observations in several ways. Most importantly, we show that the puMPS Bloch-state ansatz produces accurate excited states also in nonintegrable models, in particular the O'Brien-Fendley model and the ANNNI model \eqref{eq:H_ANNNI}, over a wide range of parameters for which the models remain critical.
As detailed above, we also check the fidelities of a large number of low-energy states in the Ising model, finding that all states are accurately reproduced. Finally, we have also tested the validity of the ansatz in the 3-state and 4-state Potts models.

\subsection{Computation of matrix elements of Virasoro generators}
The lattice Virasoro generators are constructed as Fourier modes of the Hamiltonian,
\begin{equation}
L_n+\bar{L}_{-n} \sim H_n \equiv \sum_{j=1}^N h_j e^{ij2\pi n/N}.
\end{equation}
The zero mode $H_0$ is the Hamiltonian itself, while $H_n=H^\dagger_{-n}$ are ladder operators that connect different states in the same conformal tower. For the purpose of identifying conformal towers, it suffices to compute the matrix elements of $H_n$ in the low-energy eigenbasis of the Hamiltonian,
\begin{align}
H_{n,\alpha\beta} &= \langle \phi_{\alpha}|H_n|\phi_{\beta}\rangle\\
 &= \langle \Phi_{p_\alpha}(\bar{B}_\alpha,\bar{A})|H_n|\Phi_{p_\beta}(B_\beta,A)\rangle, \label{Hnbasis}
\end{align}
where $\alpha$ and $\beta$ are labels for eigenstates in the low-energy subspace. $H_{n,\alpha\beta}$ is nonzero only if the momenta match:
\begin{equation}
\label{momenta_match}
p_\alpha+2\pi n/N=p_\beta.
\end{equation}
Since \eqref{Hnbasis} is obviously bilinear in the $B$ tensors, we can define the effective $H_n$ matrix $H_{n,\mu\nu}$ implicitly as
\begin{equation}
\label{Hn_matrix}
\bar{B}^{\mu}_{\alpha} H_{n,\mu\nu}(p_\alpha,p_\beta) B^{\nu}_{\beta}=H_{n,\alpha\beta},
\end{equation}
where $\mu,\nu$ are indices of the $B$ tensors. When $n=0$ and $p_\alpha=p_\beta=p$, $H_{n,\mu\nu}(p_\alpha,p_\beta)$ reduces to the effective Hamiltonian $H_{\mu\nu}(p)$ in \eqref{eff_H}. As we did for the effective Hamiltonian, we can also define the effective $H_n$ matrices with respect to $B_C$ as
\begin{equation}
\label{Hn_matrix_C}
 \bar{B}^{\mu}_{C,\alpha} H_{n,\mu\nu,C}(p_\alpha,p_\beta) B^{\nu}_{C,\beta}=H_{n,\alpha\beta},
\end{equation}
where the puMPS tangent vector is parametrized with $A_L$ and $	B_C=B\lambda$ as \eqref{puMPSTvec_C}.

The computation of effective $H_n$ matrices is also similar to that of the effective Hamiltonian. The tensor network for the effective Hamiltonian for $B_C$ in Fig.~\ref{fig:Neff_Hneff} is still applicable to the effective $H_n$ matrices for $B_C$, with the matrix product operator substituted by the one that represents the Fourier mode of the Hamiltonian. Thus, for each $n$ and each pair of momenta $(p_\alpha,p_\beta)$ satisfying \eqref{momenta_match}, the computational cost of $H_{n,\mu\nu,C}(p_\alpha,p_\beta)$ is also $\mathcal{O}(ND^6)$. Finally, we can plug into \eqref{Hn_matrix_C} the $B_C$ tensors for the tangent states to obtain the matrix elements $H_{n,\alpha\beta}$ of $H_n$ in the low-energy basis with a negligible $\mathcal{O}(D^2)$ cost.

\subsection{Identification of conformal towers}
With the help of matrix elements of ladder operators $H_n$, we may identify conformal towers following the method in \cite{milsted_extraction_2017}. Here, we propose a slightly different approach which identifies conformal towers level by level. The idea is that any descendant state is a linear combination of states that are obtained by successively acting $H_{n}\sim L_n+\bar{L}_{-n}$ $(n=\pm 1,\pm2)$ on the corresponding primary state, because other Virasoro generators can be obtained by their successive commutators according to the Virasoro algebra. Thus, we only need to compute the matrix elements of $H_n$ for $n=\pm1,\pm2$ to identify all states in the conformal tower.

The identification process goes as following. First, we identify primary states following the method in \cite{milsted_extraction_2017}, denoting the number of primary states as $n_p$. Second, we create a matrix $C_{\alpha l}$, where $\alpha$ labels eigenstates and $l=1,2\cdots n_p$ labels the conformal tower. The matrix is initialized to have all zero entries. Third, set $C_{\alpha(l)l}=1,\,\forall l$, where $\alpha(l)$ is the label of the $l$th primary state. This assigns each primary state to its own conformal tower. Then, for each non-primary state labelled by $\alpha$, in order of ascending energy, we update
\begin{equation}
C_{\alpha l} \leftarrow \sum_{n=-2}^{+2}\sum_{E_\beta<E_\alpha} |H_{n,\alpha\beta}|^2 C_{\beta l},
\end{equation}
normalizing the vector $C_{\alpha l}$ after each update,
\begin{equation}
C_{\alpha l} \leftarrow C_{\alpha l} \left/\sqrt{\sum_{l=1}^{n_p}C^2_{\alpha l}}\right..
\end{equation}
 Finally, for each state labelled by $\alpha$ we determine which conformal tower it belongs to according to which component in $C_{\alpha l}$ is the largest.

In the ideal case where the system size is taken to infinity, $C_{\alpha l}$ can only pick up contributions from states within the same conformal tower with scaling dimension $1$ or $2$ less than that of the state $|\Phi_\alpha\rangle$. Then $C_{\alpha l}$ should have only one nonzero entry normalized to $1$ for each $\alpha$, according to which conformal tower it belongs to. In practice, due to finite size effects induced by irrelevant perturbations, $C_{\alpha l}$ has $n_p-1$ possibly nonzero entries which go to zero with increasing system size $N$, so that only one entry remains $O(1)$ for sufficiently large system size $N$.

Typically, the largest component of $C_{\alpha l}$ decays with energy due to increasingly strong finite-size effects. However, for the Ising model, we find that the finite-size effects cause few problems in conformal tower identification for the low-lying spectrum, as long as the bond dimension of MPS is large enough to represent eigenstates accurately. We present exemplary results in Fig.~\ref{fig:ising_spec}. This is in accordance with \cite{milsted_extraction_2017}, and results from the fact that finite-size perturbation for the Ising model comes from operators in the identity tower. 

\subsection{Extrapolation of conformal data}

\begin{figure*}[h]
  \includegraphics[width=0.4\linewidth]{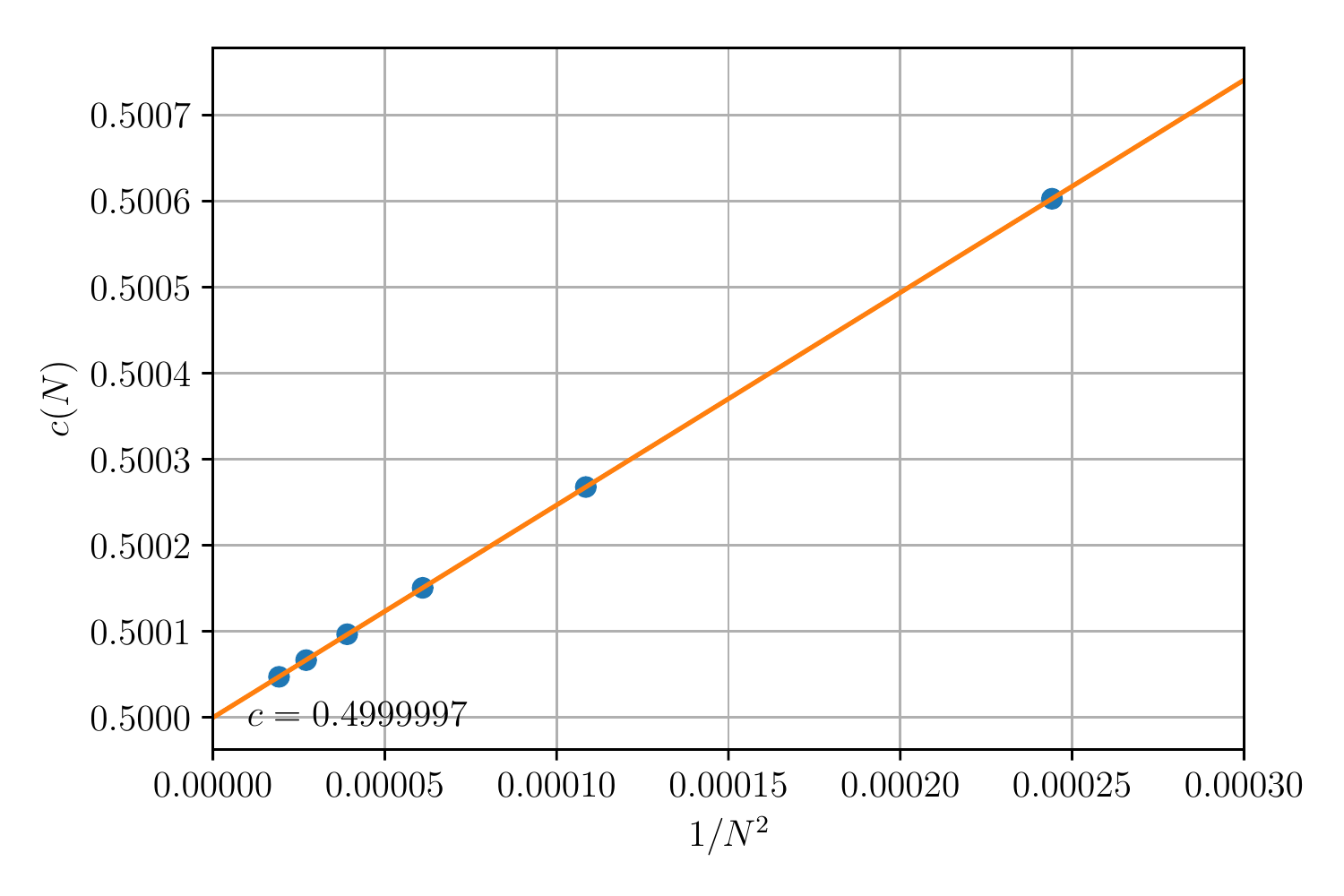}
  \includegraphics[width=0.4\linewidth]{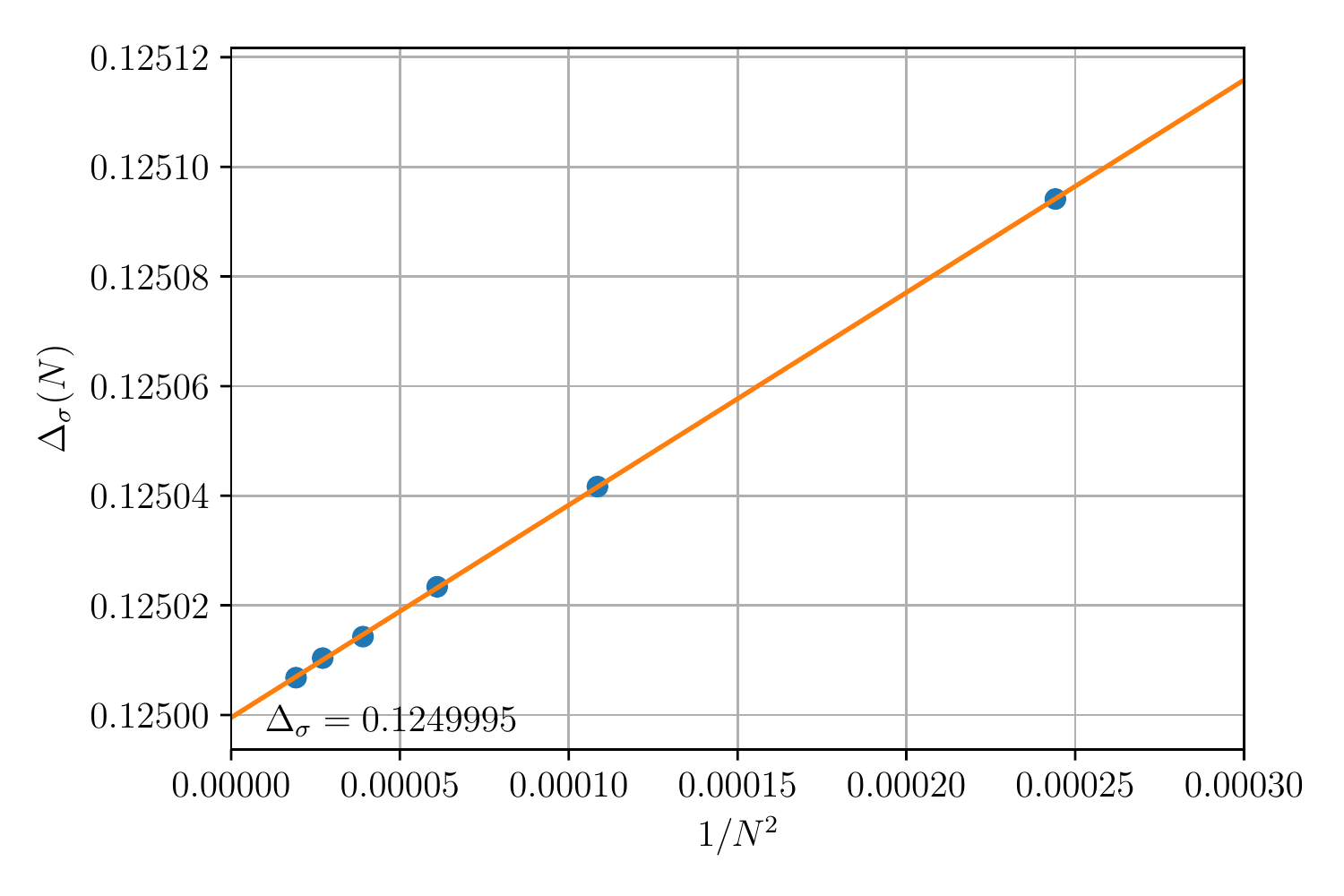}
  \includegraphics[width=0.4\linewidth]{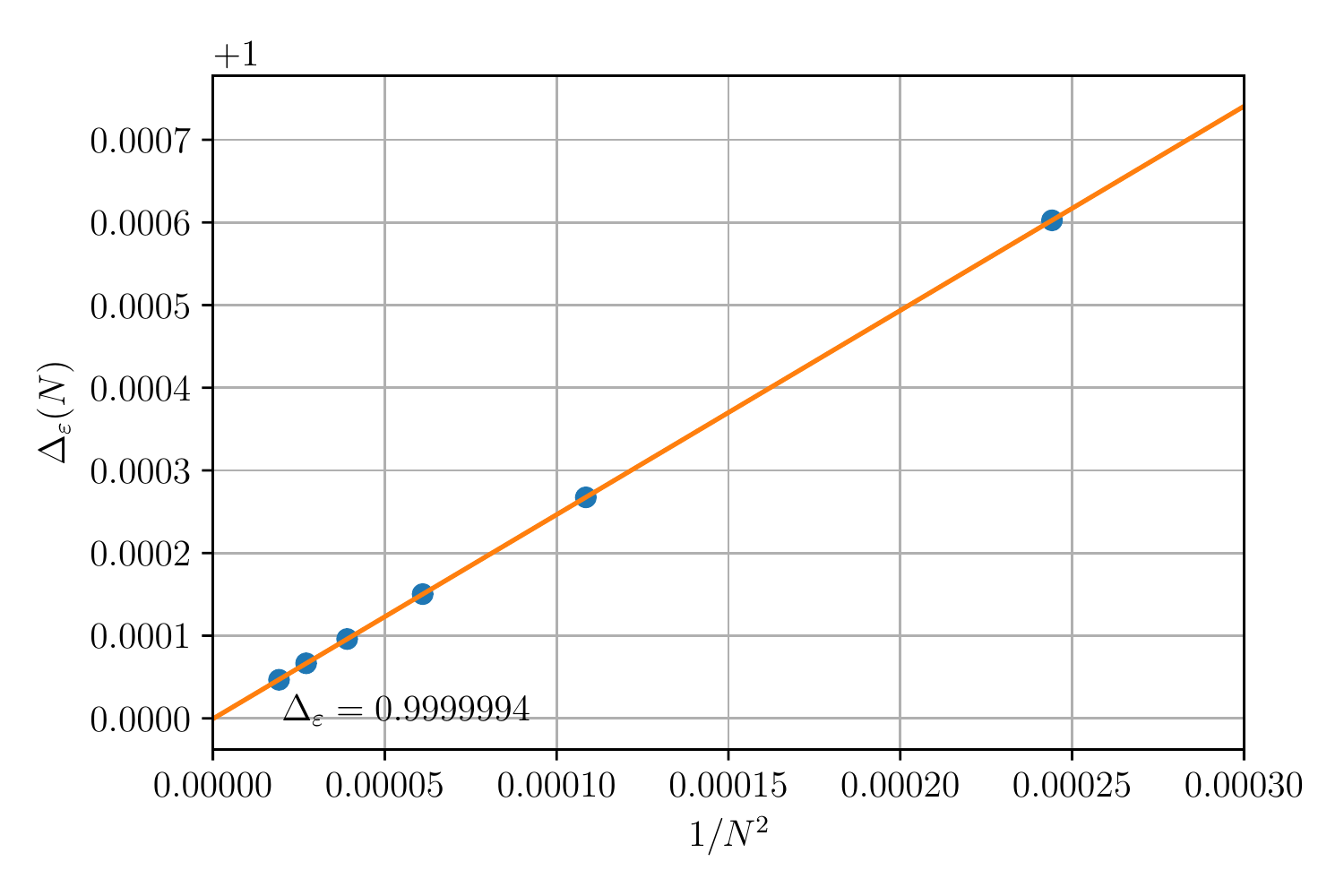}
  \includegraphics[width=0.4\linewidth]{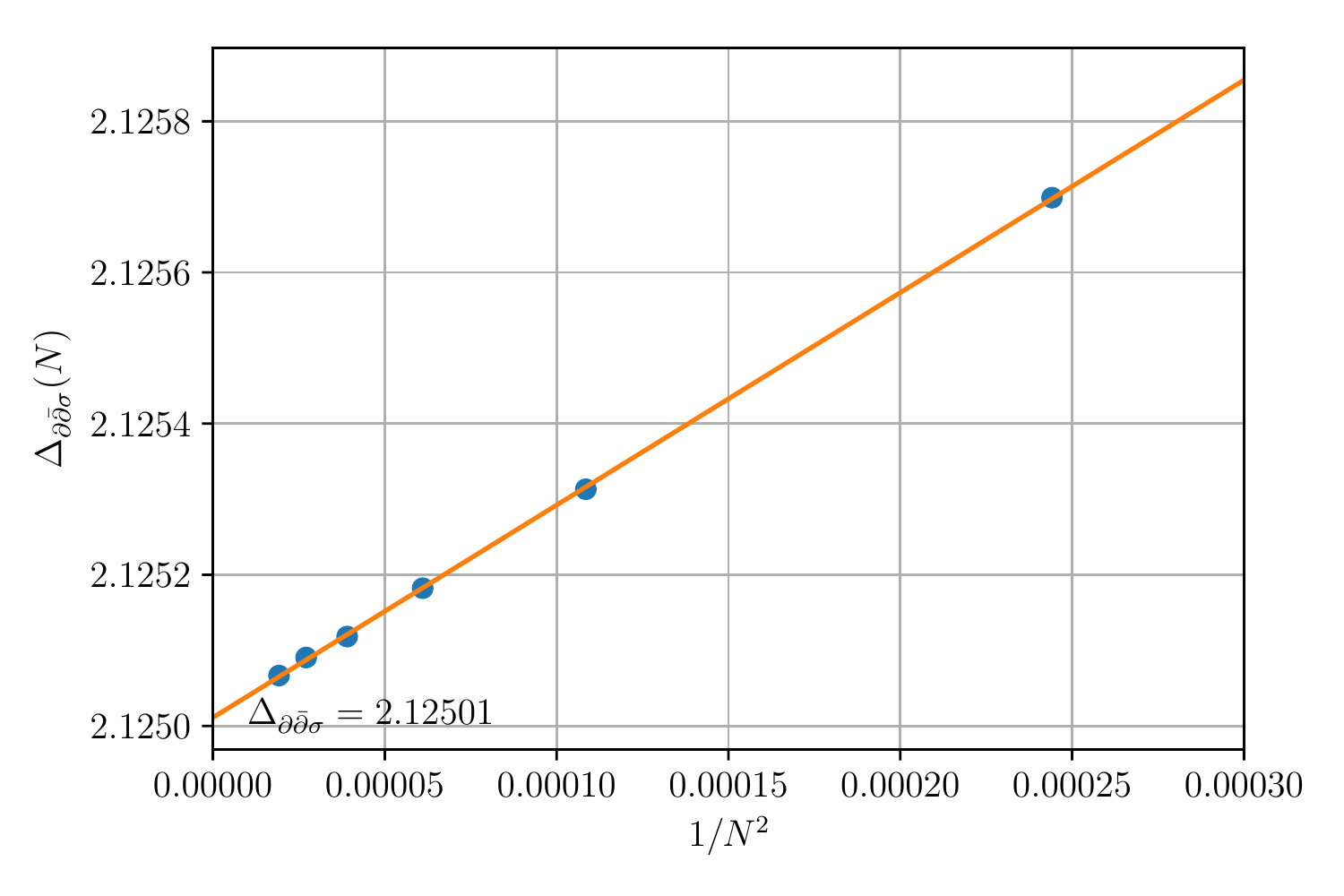}
  \includegraphics[width=0.4\linewidth]{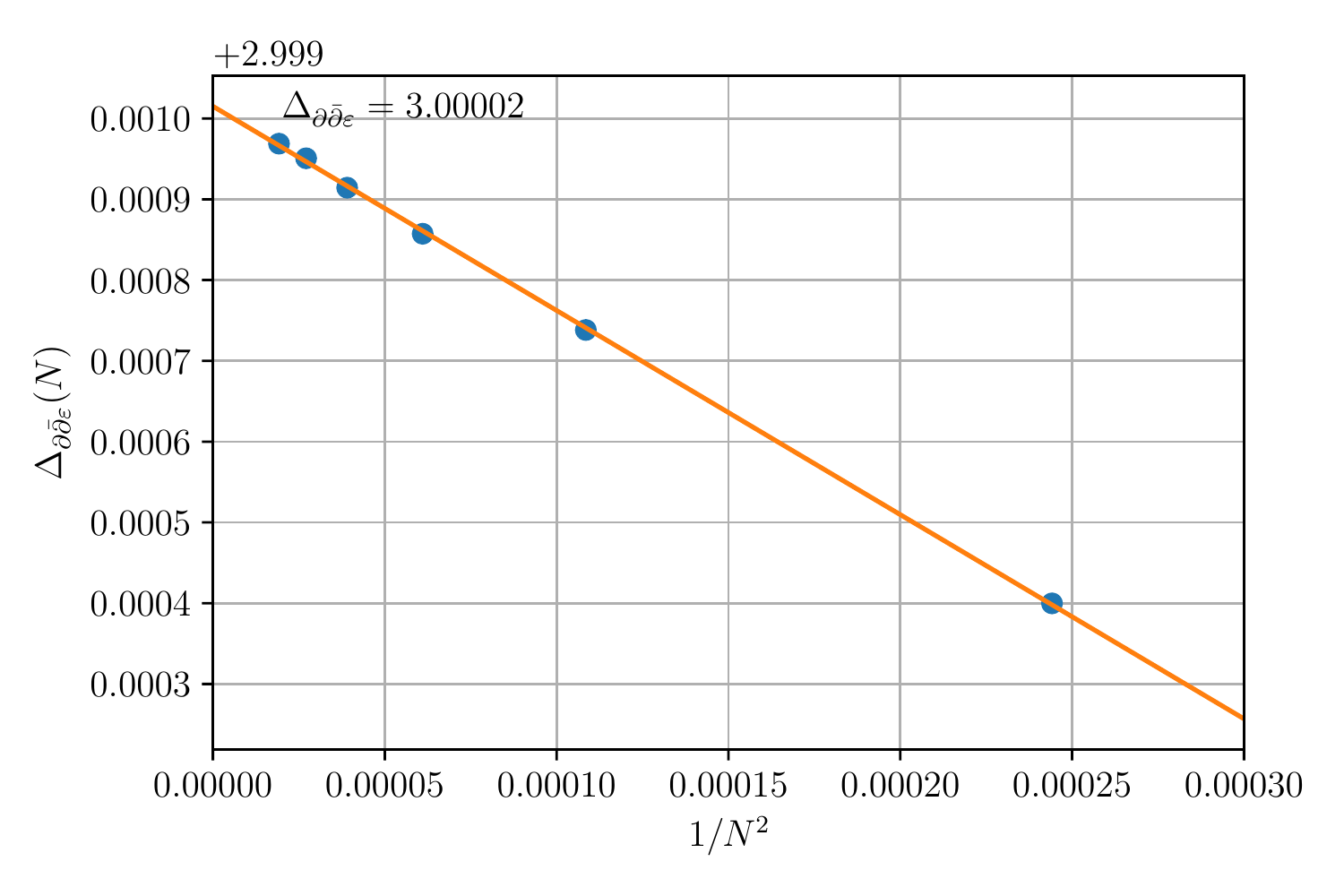}
  \includegraphics[width=0.4\linewidth]{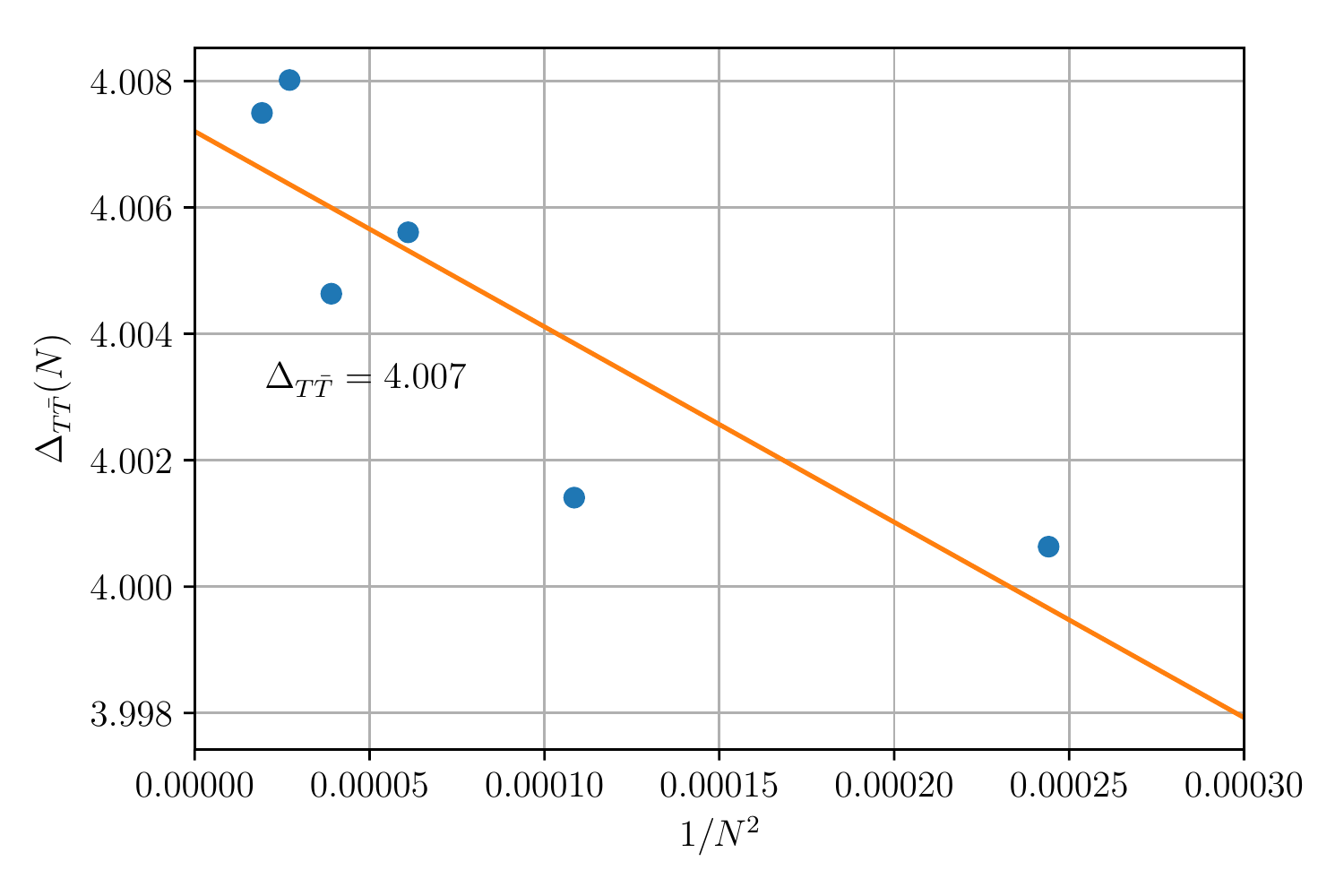}  
  \caption{\label{fig:extrapolation} Extrapolation of scaling dimensions for primary states and central charge for the Ising CFT with finite size simulations of the Ising model. Data points include $N=64,96,128,160,192,228$ with bond dimension $D=28,34,38,42,45,49$ respectively. The $T\bar{T}$ state suffers from significant finite $D$ effects for large systems with moderate bond dimensions.}
\end{figure*}

\begin{figure*}[h]
  \includegraphics[width=0.4\linewidth]{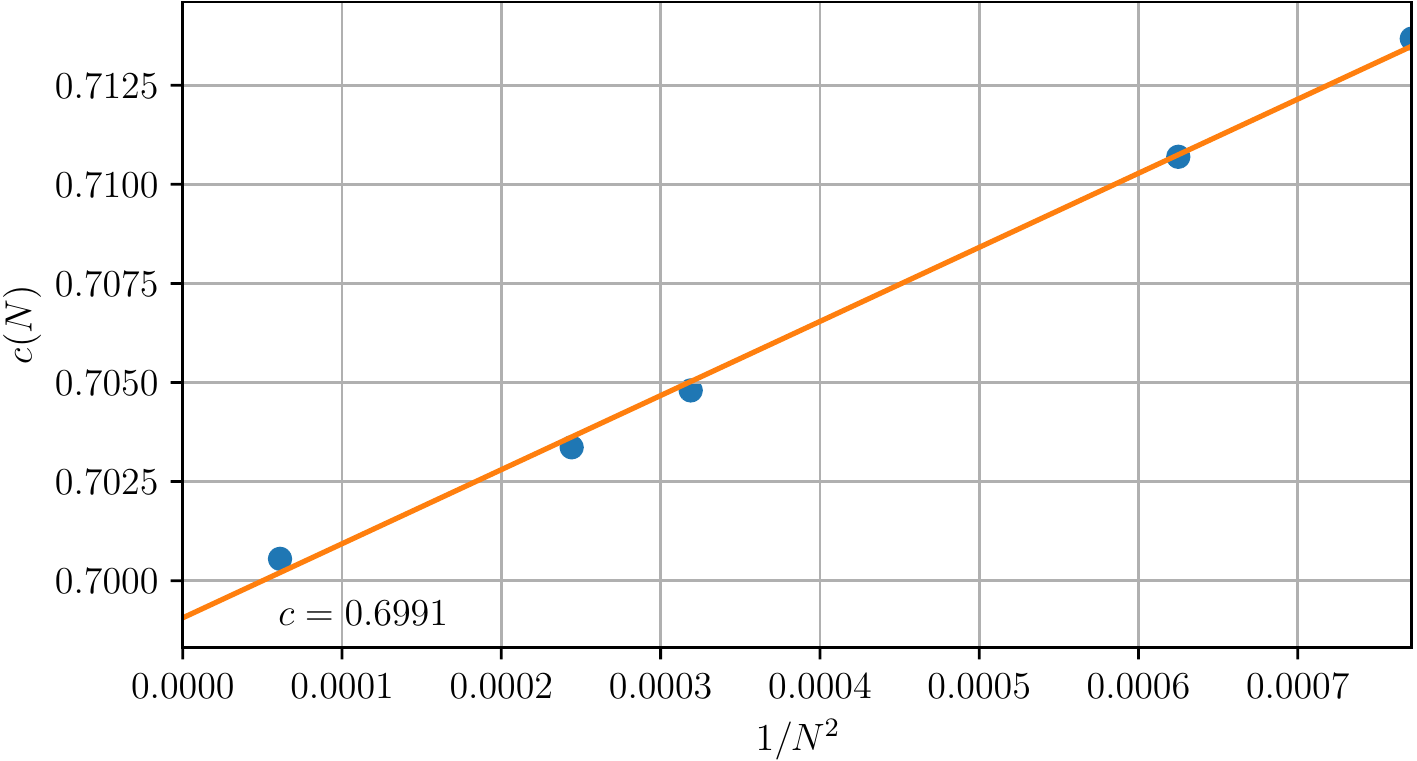} \hspace{0.4cm}
  \includegraphics[width=0.4\linewidth]{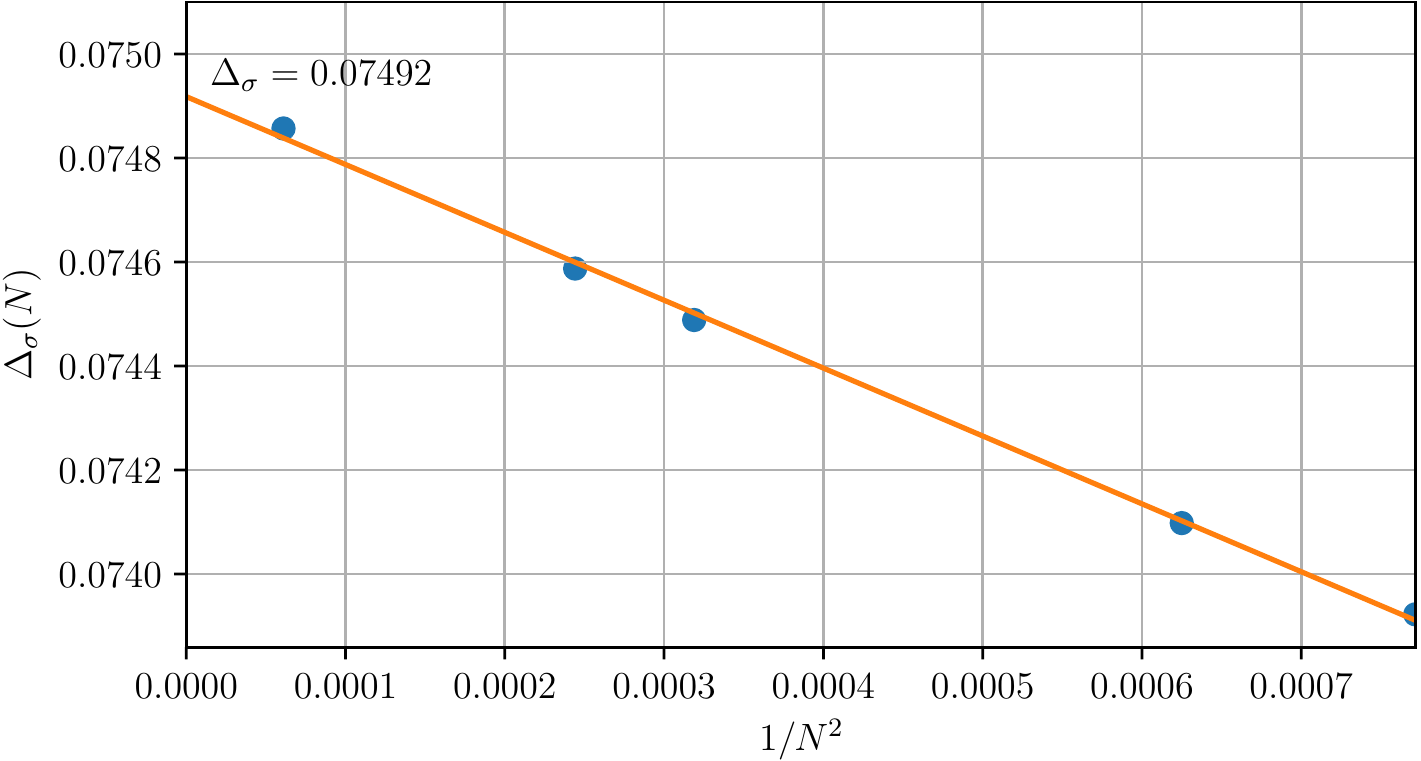}\\ \vspace{0.4cm}
  \includegraphics[width=0.4\linewidth]{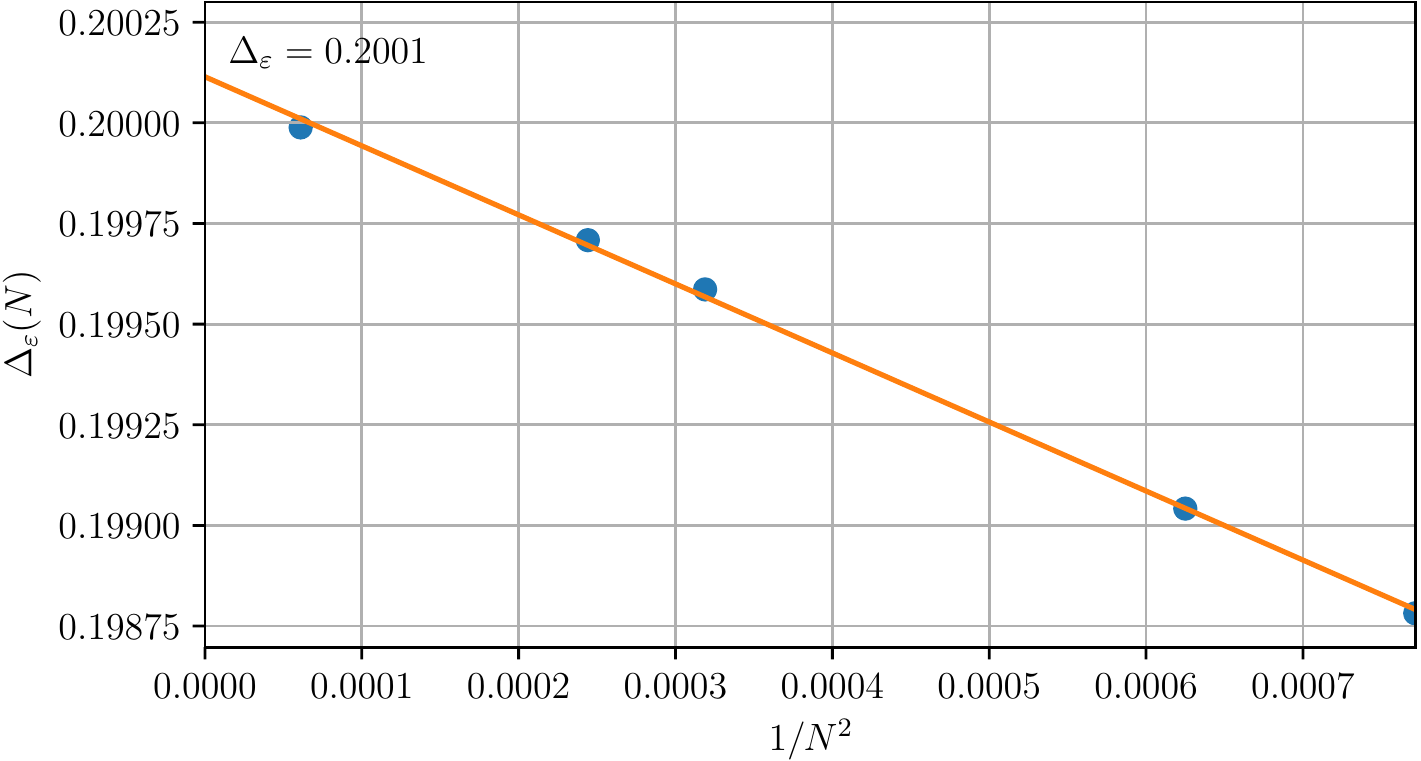} \hspace{0.4cm}
  \includegraphics[width=0.4\linewidth]{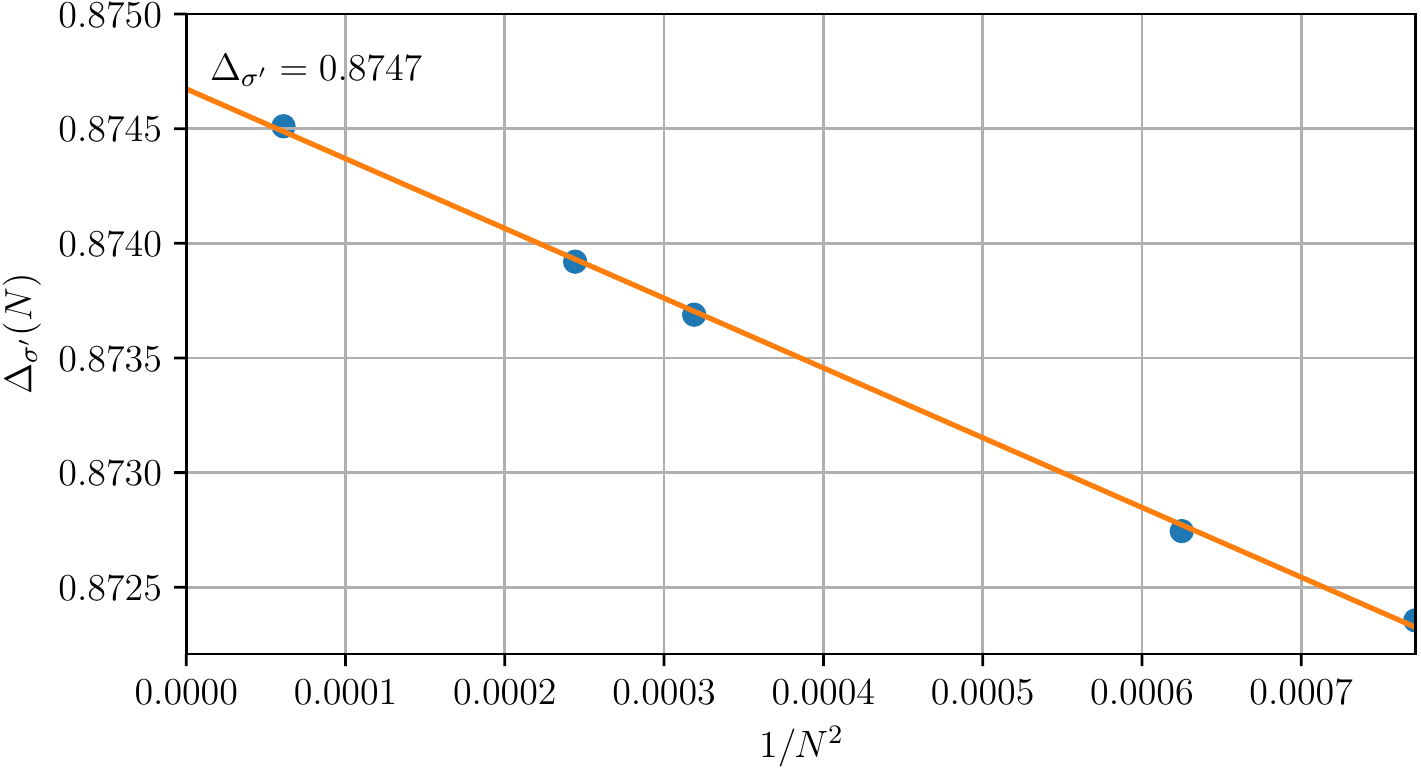}\\ \vspace{0.4cm}
  \includegraphics[width=0.4\linewidth]{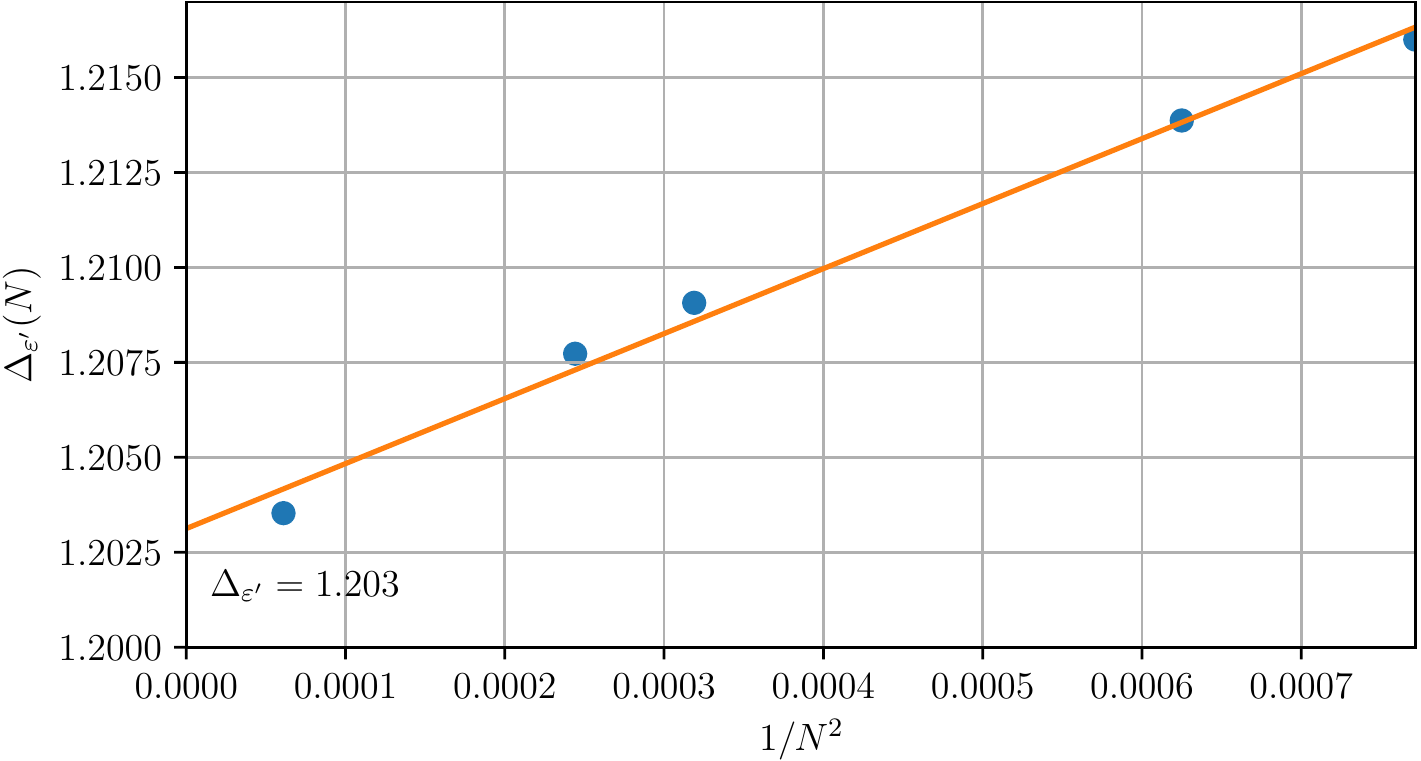} \hspace{0.4cm}
  \includegraphics[width=0.4\linewidth]{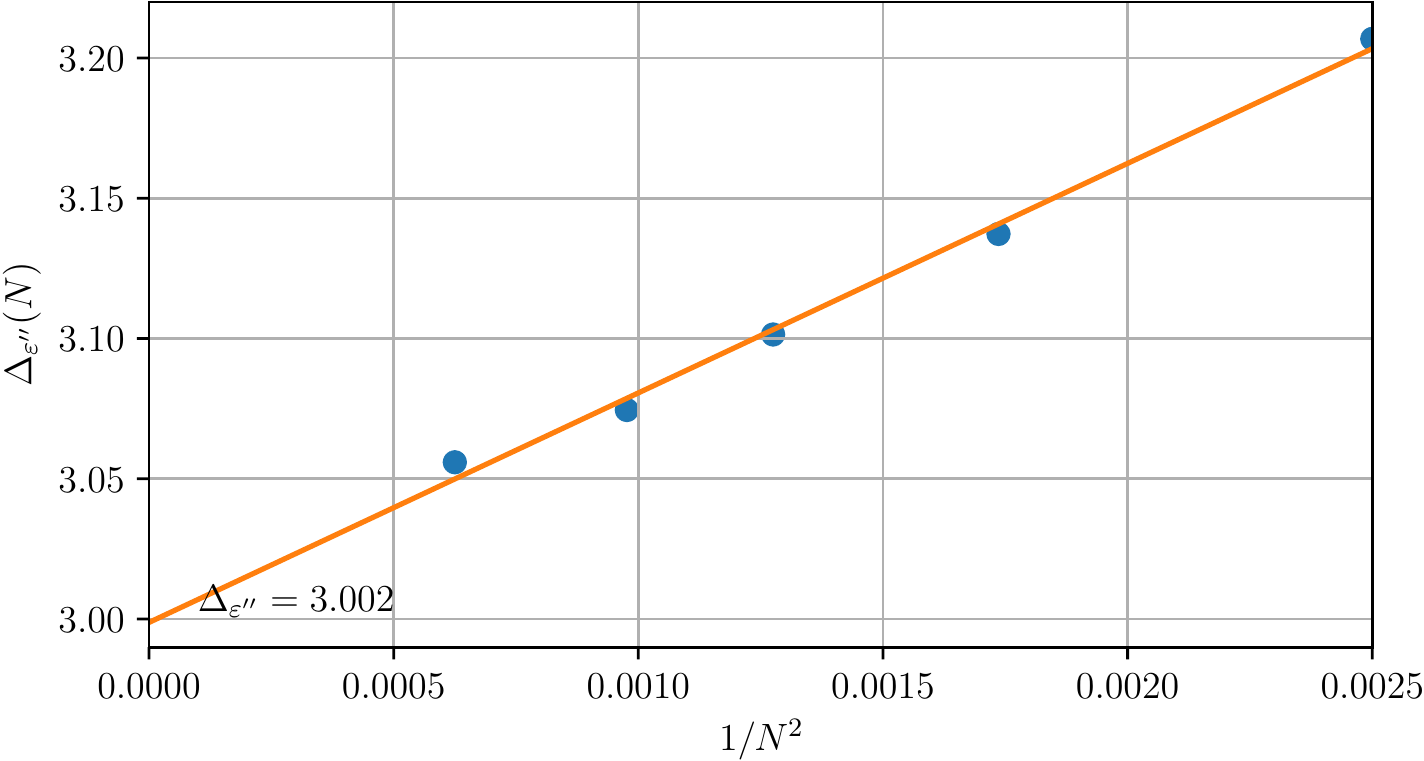}  
  \caption{\label{fig:extrapolation_TCI} Extrapolation of scaling dimensions for primary states and central charge for the Tri-Critical Ising CFT, with finite size simulations of the OF model near the TCI point. Data points include $N=36,40,56,64,128$ with bond dimensions $D=28,32,32,32,44$ respectively, except for $\Delta_{\varepsilon''}$, which uses $N=20,24,28,32,40$ with $D=24,28,28,32,32$. We chose system sizes to avoid severe corrections due to finite bond-dimension effects, which have a stronger effect on higher-energy excitations, and in order to remain in a regime where the scaling is apparently dominated by an irrelevant operator with $\Delta=4$.}
\end{figure*}

As mentioned in the main text, scaling dimensions $\Delta_\alpha$ and conformal spin $S_\alpha$ are extracted from the energy-momentum spectrum, and the central charge $c$ is extracted from the matrix element of $H_2$. There are non-universal finite-size corrections to $\Delta_\alpha$ and $c$, which depend on the particular lattice realization of CFT. By collecting $\Delta_\alpha$ and $c$ for different system sizes and extrapolating to the thermodynamic limit, we can obtain conformal data with higher accuracy.

Physical quantities in critical systems usually exhibit power law scaling. In general, we can relate the finite-size conformal data $\Delta_\alpha(N),c(N)$ to their thermodynamic values $\Delta_\alpha,c$ by
\begin{align}
\Delta_\alpha(N) &= \Delta_\alpha+\frac{b_\alpha}{N^{x_\alpha}}+o(N^{-x_\alpha}) \\
c(N) &= c+\frac{b_c}{N^{x_c}}+o(N^{-x_c}),
\end{align}
where $o()$ terms stand for higher order non-universal corrections. For the Ising model, $x_\alpha=x_c=2$ due to the presence of perturbations of operators with scaling dimension $\Delta=4$. Thus we can linearly fit finite size conformal data with $N^{-2}$ to extrapolate the thermodynamic quantities, see Fig.~\ref{fig:extrapolation}. Since we can only approximately locate the TCI point of the O'Brien-Fendley (OF) model, finite-size data obtained from the model is affected by both irrelevant and relevant corrections. The latter will eventually destroy the universal scaling of quantities as the system size increases. However, if the tuning to the TCI point is sufficiently good, we expect the universal scaling, with corrections due to irrelevant operators, to dominate at smaller system sizes. Indeed, we observe scaling compatible with a dominant contribution from a $\Delta=4$ operator, as for the Ising model, and perform extrapolations on that basis in Fig.~\ref{fig:extrapolation_TCI}. We also found some evidence for a contribution from a $\Delta=3$ operator (likely $\epsilon''$), at larger system sizes. We also note here that convergence of the puMPS variational ground state for the OF model took significantly more iterations of the energy minimization algorithm (described above) near the TCI point $\lambda = 0.428$ than at the Ising point $\lambda = 0$. For a test case of $N=30$, $D=20$, the algorithm took roughly 10 times longer to converge at the TCI point than at the Ising point, despite similar costs per iteration.

The extrapolation makes sense only if finite-size effects dominate, i.e.\ when finite bond dimension has a negligible effect. According to Fig.~\ref{fig:fidelity}, fidelity of optimized puMPS Bloch states with exact eigenstates increases with growing bond dimension $D$. Thus, we have to go to a sufficiently large bond dimension $D$ to lower the finite bond dimension errors. On the other hand, finite $D$ effects become stronger for higher excited states. Thus there is always an eigenstate for which finite $D$ errors become more significant than finite size effect. For example, in Fig.~\ref{fig:extrapolation}, with the chosen bond dimensions for different system sizes, finite $D$ affects the $T\bar{T}$ state far more seriously than other lower-energy states such that the extrapolation cannot produce an accurate estimate.

\subsection{Comparison with other methods}
Here we present a comparison of the conformal data extracted from the critical Ising model using puMPS, with similar data (from the literature) computed using other tensor-network methods. We compare our results with finite entanglement scaling (FES) based on infinite matrix product states \cite{stojevic_2015}, the tensor renormalization group (TRG) \cite{levin_tensor_2007}, and tensor network renormalization (TNR) \cite{evenbly_tensor_2015}. These methods derive conformal data from different quantities: In \cite{stojevic_2015}, scaling dimensions are computed from the decay exponents of two-point correlation functions of scaling operators on the lattice and the central charge is extracted from the scaling of the entanglement entropy with the bond dimension. In \cite{evenbly_tensor_2015}, which presents data for TRG as well as TNR, the central charge and scaling dimensions are extracted from eigenvalues of a coarse-grained transfer matrix. 

\begin{table}[h]
  \begin{center}
    \footnotesize
    \begin{tabular}{llllll}
\toprule
 & exact & puMPS  & FES &TRG  & TNR     \\
\midrule
$c$ & 0.5  & 0.4999997 & 0.496 & 0.49982 & 0.50001 \\
$\Delta_{\sigma}$ & 0.125 & 0.1249995 & 0.1246 & 0.12498 & 0.1250004  \\
$\Delta_\varepsilon$ & 1 & 0.9999994 & 0.998 & 1.00055 & 1.00009 \\
$\Delta_{\partial \sigma}$ & 1.125 & 1.1249994 & \multirow{2}{*}{1.12485} & 1.12615* & 1.12492* \\
$\Delta_{\overline{\partial} \sigma}$ & 1.125 & 1.1249994 &  & 1.12635* & 1.12510* \\
$\Delta_{\partial\varepsilon}$ & 2 & 1.9999998 & \multirow{2}{*}{1.9985} & 2.00243* & 1.99922* \\
$\Delta_{\overline{\partial}\varepsilon}$ & 2 & 1.9999998 &  & 2.00579* & 1.99986* \\
$\Delta_{T}$ & 2 & 2**              & -------- & 2.00750* & 2.00006* \\
$\Delta_{\overline{T}}$ & 2 & 2**               & -------- & 2.01061* & 2.00168* \\
\bottomrule
  \end{tabular}
\end{center}
\caption{\label{tbl:comp} Central charge and selected scaling dimensions extracted from the critical Ising model, comparing the puMPS techniques we employ to finite entanglement scaling (FES) with infinite matrix product states \cite{stojevic_2015}, the tensor renormalization group (TRG) \cite{levin_tensor_2007}, and tensor network renormalization (TNR) \cite{evenbly_tensor_2015}. Note that, for FES, the scaling dimensions $\approx 1.125$ correspond to the \emph{spatial}-derivative operators $\partial_x \sigma$ and $\partial_x \epsilon$ (denoted $d\sigma$ and $d\varepsilon$ in \cite{stojevic_2015}), which are mixtures of $\partial\sigma$, $\overline{\partial}\sigma$ and $\partial \varepsilon$, $\overline{\partial}\varepsilon$, respectively. To indicate this, we have placed these values \emph{between} rows. Also, values marked with $*$ were not assigned to particular CFT operators in \cite{evenbly_tensor_2015} so we have simply listed them in ascending order. Finally, in the puMPS data, the values for $\Delta_T$ and $\Delta_{\bar{T}}$ (marked with $**$) are exact because these scaling dimensions were used to fix the overall normalization.
The bond dimensions used were $28\leq D\leq 49$ for puMPS, $32\leq D\leq 64$ for FES, 64 for TRG, and 24 for TNR.}
\end{table}

Examining the results, shown in Table~\ref{tbl:comp}, we find that the accuracy of conformal data extracted from puMPS is consistently better than for the other methods. We also remark that, in the case of puMPS, a \emph{complete} set of scaling dimensions (and conformal spins) can, in principle, be extracted systematically, together with the \emph{identity} of the scaling operator corresponding to each scaling dimension. This is not the case in \cite{stojevic_2015}, where extracting scaling dimensions requires knowledge of lattice versions of each scaling operator of interest, or in TRG/TNR, where the scaling operator corresponding to the computed scaling dimensions was not identified.

\subsection{Lattice Virasoro generators and the O'Brien-Fendley model}

One may improve on the procedures of \cite{milsted_extraction_2017} for the O'Brien-Fendley model  \cite{obrien_lattice_2018} studied in the main text, whose Hamiltonian we reproduce here:
\begin{multline}
  H = \sum_{j=1}^N \Big[ -\sigma^Z_j \sigma^Z_{j+1} - \sigma^X_j + \\
   \lambda \left(\sigma^X_j \sigma^Z_{j+1} \sigma^Z_{j+2} + \sigma^Z_j \sigma^Z_{j+1} \sigma^X_{j+2} \right)\Big],
\end{multline}
To construct the lattice Virasoro generator $H_n \sim L_n + \overline{L}_{-n}$ according to \cite{milsted_extraction_2017}, phases are assigned to terms in $H$ according to the midpoints of each term, resulting in the same phase for both components of the $\lambda$ term. This is, however, incompatible with covariance of $H_n$ under duality, which would demand that the location of these two terms differ by half a lattice site, so that their phases differ by $e^{\ic n \pi/N}$. Using the rule of \cite{milsted_extraction_2017} thus results in contamination of $H_n$ with some amount of a duality-odd operator, leading to additional finite-size corrections in e.g.\ conformal-tower identification.

To obtain more accurate results, we therefore use some extra knowledge of $H$, namely the expression of $\sigma^X_j \sigma^Z_{j+1} \sigma^Z_{j+2}$ and $\sigma^Z_j \sigma^Z_{j+1} \sigma^X_{j+2}$ in terms of Majorana fermion operators under an appropriate Jordan-Wigner transformation \cite{obrien_lattice_2018}, assigning phases to the $\lambda$ terms in $H$ according to the midpoints of their equivalent Majorana fermion operators. This results in
\begin{multline}
  H_n \equiv \frac{N}{2\pi} \sum_{j=1}^N \Big[-e^{\ic j n \frac{2\pi}{N}} \sigma^X_j - e^{\ic (j+\frac{1}{2}) n \frac{2\pi}{N}}  \sigma^Z_{j} \sigma^Z_{j+1} + \\
  \lambda\Big(e^{\ic (j+\frac{3}{4}) n \frac{2\pi}{N}} \; \sigma^X_j \sigma^Z_{j+1} \sigma^Z_{j+2} + \\
  \qquad e^{\ic (j+\frac{5}{4}) n \frac{2\pi}{N}}\; \sigma^Z_j \sigma^Z_{j+1} \sigma^X_{j+2} \Big) \Big].
\end{multline}

\subsection{Spectral flow comparison with integrable field theory}

We have shown in the main text that the energy gaps of a critical lattice model on a circle can be computed with puMPS Bloch states. In particular, we investigated the O'Brien-Fendley (OF) model
\begin{multline}
\label{lattice_H_TCI}
  H = \sum_{j=1}^N \Big[-\sigma^Z_j \sigma^Z_{j+1}-\sigma^X_j + \\
  (\lambda_{\TCI} - \delta)\left(\sigma^X_j \sigma^Z_{j+1} \sigma^Z_{j+2}+\sigma^Z_j \sigma^Z_{j+1} \sigma^X_{j+2}\right)\Big], 
\end{multline}
which is in the Tricritical Ising (TCI) universality class for $\delta = 0$. Turning on $\delta$ introduces a relevant perturbation that, for $\delta > 0$, changes the universality class to that of the Ising CFT. The dominant contribution to the relevant perturbation corresponds to the TCI primary operator $\varepsilon'$ ($\phi_{1,3}$ in the Kac table \cite{friedan_conformal_1984}).

In the main text, we studied the spectal flow, with the system size, of the energy gaps in this model for $\delta > 0$. We additionally compared the flow of the first gap with a result \cite{klassen_spectral_1992} for an integrable quantum field theory. In \cite{klassen_spectral_1992} the authors considered the spectral flow, with the system size, of the first energy gap of a quantum field theory Hamiltonian of the form
\begin{equation}
\label{FT_H_TCI}
H^{\QFT}=H^{\TCI}+\tilde{\delta} \int_0^L dx \, \varepsilon'(x),
\end{equation}
where $H^{\TCI}$ and $\varepsilon'(x)$ are QFT realizations of the TCI CFT Hamiltonian and the $\varepsilon'$ primary field, respectively, so that $H^\QFT$ can be thought of as the continuum theory corresponding to $H$, with $\tilde \delta \sim \delta$. We require that $H^{\TCI}$ is normalized such that the speed of light is $1$ and that $\varepsilon'(x)$ is normalized such that
\begin{equation}
\langle \varepsilon' |  \varepsilon'(x) |0\rangle=\left(\frac{2\pi}{L}\right)^{\Delta_{\varepsilon'}}.
\end{equation}
The authors of \cite{klassen_spectral_1992} express the first energy gap of $H^\QFT$ as
\begin{equation}
E^{\QFT}_{\sigma}-E^{\QFT}_0=\frac{2\pi}{L}e(r),
\end{equation}
where
\begin{equation}
r=L\left(\frac{\tilde{\delta}}{\kappa}\right)^{5/4}.
\end{equation}
The dimensionless quantity $e(r)$ is determined by solving a set of equations, proposed in \cite{klassen_spectral_1992} and conjectured to produce the correct result for the gap. The constant $\kappa\approx 0.148696$ is needed to match the predictions with those of conformal perturbation theory (e.g.\ the RG flow from TCI CFT and Ising CFT suggests that $e(0^{+})=\Delta^\TCI_\sigma=3/40$ and $e(+\infty)=\Delta_\sigma^{\textsl{\tiny Ising}}=1/8$).

To compare (\ref{lattice_H_TCI}) with ({\ref{FT_H_TCI}}), we have to be careful in determining the correct normalization of $H$. First, we identify $L$ in ({\ref{FT_H_TCI}}) with the system size $N$ in (\ref{lattice_H_TCI}). Second, at the TCI point $\delta=0$, we multiply the Hamiltonian by $\eta^{*}$ such that the speed of light is $1$, i.e., as $N\rightarrow \infty$, 
\begin{equation}
\eta^{*}\times(E^{*}_T(N)-E^{*}_0(N))=\frac{2\pi}{N}\times 2,
\end{equation}
where $E^*_0$ and $E^*_T$ are the energies of the ground state and the stress-tensor states and $*$ indicates the TCI point. Third, the operator appearing in the $\delta$ term is related to the CFT operator $\varepsilon'$ as
\begin{equation}
\sigma^X_j \sigma^Z_{j+1} \sigma^Z_{j+2}+\sigma^Z_j \sigma^Z_{j+1} \sigma^X_{j+2} \sim C_{\varepsilon'} \varepsilon'(x) + \cdots,
\end{equation}
where $\cdots$ represents irrelevant operators. The coefficient $C_{\varepsilon'}$ can then be estimated using
\begin{equation}
|\langle \varepsilon'^{*} |\sigma^X_j \sigma^Z_{j+1} \sigma^Z_{j+2}+\sigma^Z_j \sigma^Z_{j+1} \sigma^X_{j+2}|0^{*}\rangle|=C_{\varepsilon'}\left(\frac{2\pi}{N}\right)^{\Delta_{\varepsilon'}},
\end{equation}
computed for sufficiently large $N$, where the states are eigenstates of $H$ at the TCI point. We can then relate the lattice and QFT parameters as
\begin{equation}
\tilde{\delta}= C_{\varepsilon'}\eta^{*} \delta.
\end{equation}
Using this identification, we can also relate $N$ and $r$ as
\begin{equation}
N=r\left(\frac{\tilde{\delta}}{\kappa}\right)^{-5/4}.
\end{equation}
Carrying out this procedure, we approximately extract $C_{\varepsilon'}\approx 0.8235$ and $\eta^{*}\approx 0.6147 $ with the numerical data at $N=128$ and $D=44$ at the approximate TCI point $\lambda_{TCI}\approx 0.428$. In the main text, we study the spectral flow of $H$ at $\delta=0.028$, resulting in $\tilde \delta \approx 0.0142$.

In the main text, we plotted gaps as a function of $N$ in terms of
\begin{equation}
\Delta_\alpha(N)=2\frac{E_\alpha(N)-E_0(N)}{E_T(N)-E_0(N)},
\end{equation}
where $\alpha$ refers to an excited state, which scales all gaps such that, for the stress-tensor state $|T\rangle$, we have $\Delta_T = 2$ for all $N$. This $N$-dependent scaling of energies has the advantage that, at small system sizes for small $\delta > 0$, the $\Delta_\alpha$ for the lower-energy gaps are close to TCI scaling dimensions. However, for comparison with the gaps determined in \cite{klassen_spectral_1992}, we require an $N$-independent scale factor.

To do this, we scale $H$ such that the speed of light is $1$ in the IR (as $N\rightarrow\infty$), using a normalization constant $\eta$ given by
\begin{equation}
\eta=\lim_{N\rightarrow\infty} \frac{2\pi}{N}\frac{2}{E_T(N)-E_0(N)},
\end{equation}
where the limit is taken numerically by a linear extrapolation with $1/N\rightarrow 0$. Note that the correct normalization constant $\eta$ of the perturbed Hamiltonian is not the same as $\eta^{*}$, since $\delta$ is not infinitesimal. We then compute the scaled gap
\begin{equation}
e(N)=\frac{N}{2\pi}(E_\sigma(N)-E_0(N))\eta,
\end{equation}
which we compare to the QFT result $e(r)$ in the main text.

We found good agreement, demonstrating the consistency of our lattice results with the QFT conjectures of \cite{klassen_spectral_1992}. The agreement is best at large system sizes, becoming worse as the system size decreases. This is reasonable, since at smaller system sizes irrelevant perturbations become more significant and the lattice model starts to depart from the underlying field theory description.

\subsection{Flows in the ANNNI model}
\label{sec:ANNNI}

\begin{figure}[t]
  \vspace{1em}
  \includegraphics[width=\linewidth]{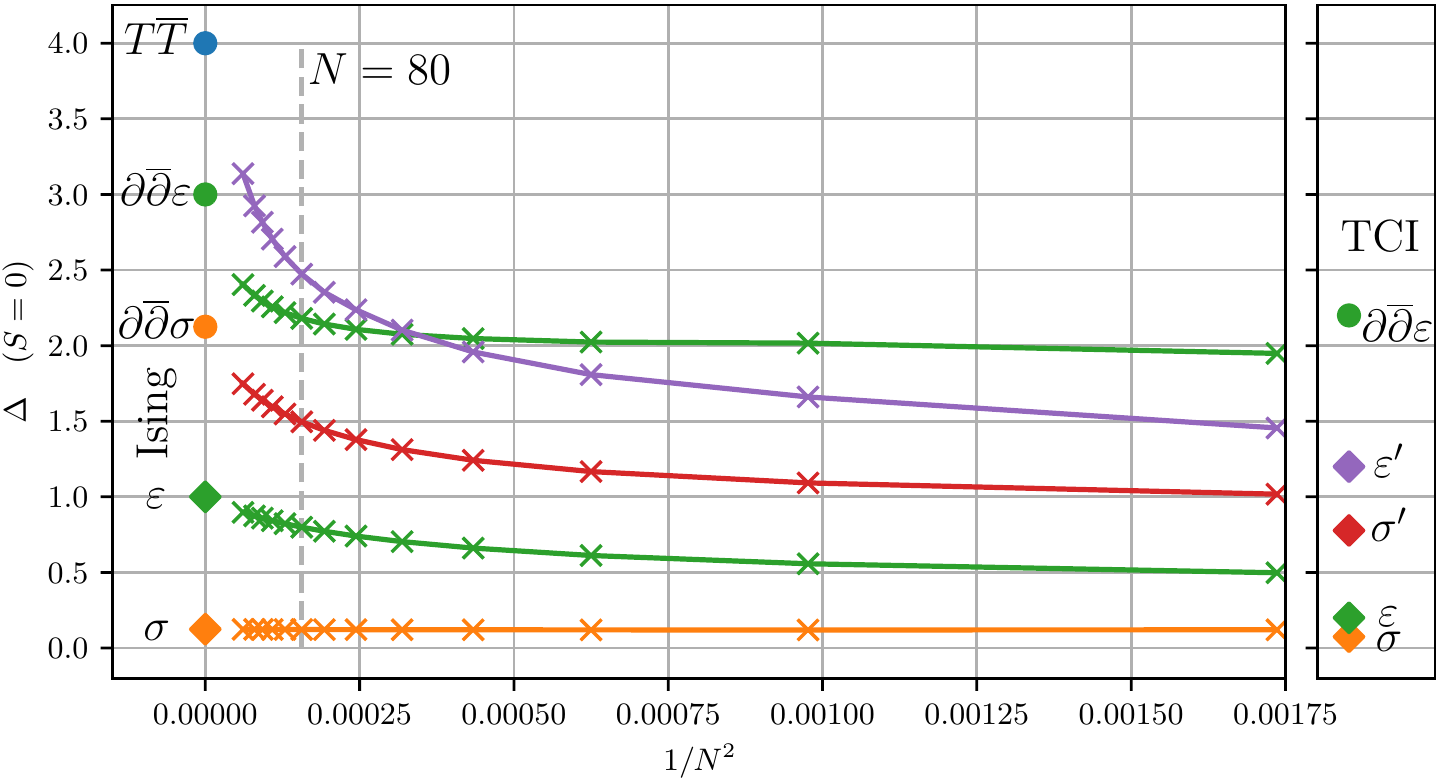}
  \caption{\label{fig:ANNNI_flow_N} Spectral RG flow of the first 5 approximate scaling dimensions (crosses), excluding $\Delta=0$, extracted from the ANNNI model at momentum zero, for $\gamma=10$, using $D \le 46$. For comparison, we also plot the exact scaling dimensions of the Ising and TCI CFTs. Note the crossover between the two largest scaling dimensions plotted, which we confirm by also tracking the $\varepsilon$-tower membership using $H_n$ matrix elements.}
\end{figure}

\begin{figure}
  \includegraphics[width=\linewidth]{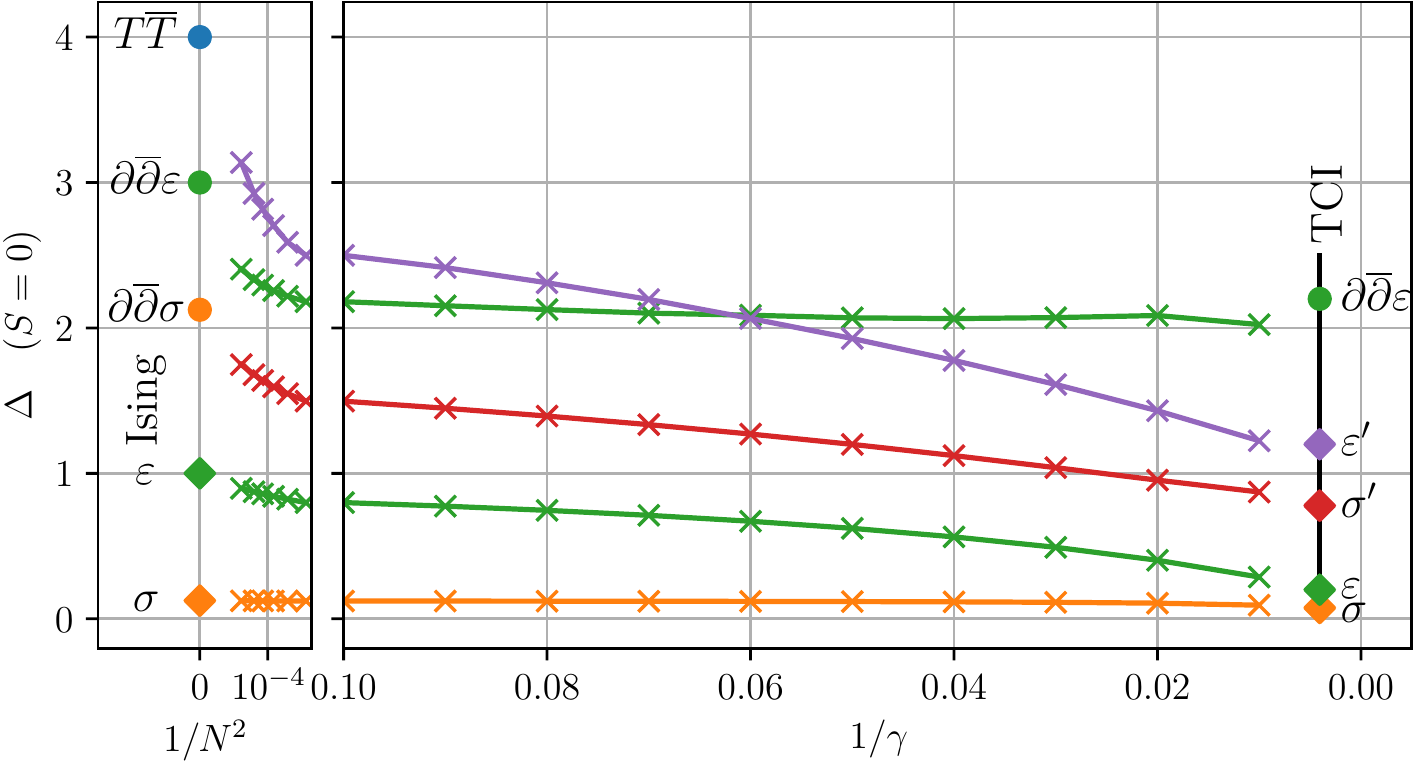}
  \caption{\label{fig:ANNNI_flow_gamma} The first 5 approximate scaling dimensions (crosses), excluding $\Delta=0$, as function of $\gamma$, extracted from the ANNNI model at momentum zero, for $N=80, D=38$. We also plot exact CFT scaling dimensions. Furthermore, we show to the left how the ``flow'' with $\gamma$ links up at $\gamma=10$ with the spectral RG flow of Fig.~\ref{fig:ANNNI_flow_N}. We confirm the crossover between the two highest-$\Delta$ curves by tracking fidelities of excited states at different $\gamma$.}
\end{figure}

In the main text we study flows of energy gaps in a model \cite{obrien_lattice_2018} with a gapless Ising phase as well as a Tricritical Ising point. Another model with these properties is the Anisotropic Next-Nearest-Neighbor Ising (ANNNI) model
\begin{equation} \label{eq:H_ANNNI}
  H = -\sum_{j=1}^N \left[ \sigma^Z_j \sigma^Z_{j+1} + \sigma^X_j + \gamma \left(\sigma^Z_j \sigma^Z_{j+2} + \sigma^X_j \sigma^X_{j+1} \right)\right],
\end{equation}
which includes the critical Ising model at $\gamma = 0$ and features a Tri-Critical Ising (TCI) point at $\gamma_\TCI \approx 247$ \cite{rahmani_emergent_2015}. It is symmetric under $\sigma^Z \rightarrow -\sigma^Z$ and is self dual for all $\gamma$ \cite{milsted_statistical_2015, rahmani_phase_2015-1}.

Note that the scale of the $\gamma$ term in \eqref{eq:H_ANNNI} at $\gamma_\TCI$ is two orders of magnitude larger than that of the remaining Hamiltonian. Compared to the O`Brien-Fendley model of the main text, this makes the ANNNI model more difficult to study numerically, as the resulting linear algebra problems involved in using puMPS techniques are relatively ill-conditioned. Nevertheless, we were able to extract an RG flow (Fig.~\ref{fig:ANNNI_flow_N}) for the ANNNI model that compares well with that of the main text. The gaps are also plotted as a function of $\gamma$ in Fig.~\ref{fig:ANNNI_flow_gamma}. Due to very slow convergence of the puMPS ground state, we had difficulty reaching the TCI point with the chosen system size and bond dimension.

\end{document}